\documentclass[a4paper,11pt]{article}
\pdfoutput=1 % if your are submitting a pdflatex (i.e. if you have
\usepackage{jheppub} % for details on the use of the package, please
\usepackage[T1]{fontenc} % if needed
\usepackage{cleveref}
\usepackage{feynmp-auto}
\usepackage{subcaption}
\usepackage{accents}
\usepackage{xspace}
\usepackage{enumitem}
\newlength{\dhatheight}
\newcommand{\doublehat}[1]{%
    \settoheight{\dhatheight}{\ensuremath{\hat{#1}}}%
    \addtolength{\dhatheight}{-0.35ex}%
    \hat{\vphantom{\rule{1pt}{\dhatheight}}%
    \smash{\hat{#1}}}}

\newcommand{\bg}{\ensuremath{\beta_g}\xspace}
\newcommand{\bgs}{\ensuremath{\beta_g^2}\xspace}
\newcommand{\MET}{\ensuremath{E_\text{T}^\text{miss}}\xspace}
\newcommand{\HT}{\ensuremath{H_\text{T}}\xspace}
\newcommand{\pT}{\ensuremath{p_\text{T}}\xspace}
\newcommand{\Nb}{\ensuremath{N_{b\text{-jet}}}\xspace}
\newcommand{\mll}{\ensuremath{m_{\ell\ell}}\xspace}
\newcommand{\ggf}{$gg$F\xspace}
\newcommand{\dphill}{\ensuremath{\Delta\phi(\ell^+,\ell^-)}\xspace}
\newcommand{\mur}{\ensuremath{\mu_\text{R}}\xspace}
\newcommand{\muf}{\ensuremath{\mu_\text{F}}\xspace}

\title{The emergence of multi-lepton anomalies at the LHC and their compatibility with new physics at the EW scale}

\author[a,1]{Stefan von Buddenbrock,\note{Corresponding author.}}
\author[a]{Alan S. Cornell,}
\author[b,c,d]{Yaquan Fang,}
\author[a,b,d]{Abdualazem Fadol Mohammed,}
\author[a]{Mukesh Kumar,}
\author[a,e]{Bruce Mellado}
\author[a]{and Kehinde G. Tomiwa}

\affiliation[a]{School of Physics and Institute for Collider Particle Physics, University of the Witwatersrand, Johannesburg, Wits 2050, South Africa}
\affiliation[b]{Institute of High Energy Physics, 19B Yuquan Road, Shijing District, Beijing 100049, China}
\affiliation[c]{Beijing Advance Sciences and Innovation Center, TechartPlaz, 30 Xueyuan Road Haidian district, Beijing, 100083, China}
\affiliation[d]{University of Chinese Academy of Sciences, 19A Yuquan Road, Shijing District, Beijing 100049, China}
\affiliation[e]{iThemba LABS, National Research Foundation, PO Box 722, Somerset West 7129, South Africa}

\emailAdd{stef.von.b@cern.ch}
\emailAdd{Alan.Cornell@wits.ac.za}
\emailAdd{fangyq@ihep.ac.cn}
\emailAdd{abdualazem.fadol.moohammed@cern.ch}
\emailAdd{mukesh.kumar@cern.ch}
\emailAdd{Bruce.Mellado@wits.ac.za}
\emailAdd{kehinde.gbenga.tomiwa@cern.ch}

\abstract{A recent study~\cite{vonBuddenbrock:2017gvy} has shown that a simplified model predicting a heavy scalar of mass 270~GeV ($H$) that decays to a Standard Model (SM) Higgs boson in association with a scalar singlet of mass 150~GeV ($S$) can accommodate several anomalous multi-lepton results in proton-proton collisions at the Large Hadron Collider (LHC).
With this in mind, the goal of this article is to provide a more formal study of a wider set of LHC results pertaining to the production of multiple leptons. 
We find that a combination of such results lead to strong discrepancies between the data and SM Monte Carlo predictions. 
These discrepancies appear in corners of the phase-space where different SM processes dominate, indicating that the potential mismodeling of a single SM process is unlikely to explain them. 
Systematic uncertainties from the prediction of SM processes evaluated with currently available tools seem unable to explain away these discrepancies. 
A combination is able to constrain the simplified model's single degree of freedom \bgs, related to the size of the Yukawa coupling of $H$ to the top quark, to a value of $2.92\pm 0.35$. 
This is in contrast to the absence of signal, where $\bg=0$.
This result is discussed in the independent contexts of both potential for new physics in the existing LHC data as well as the limitations of our current understanding of the SM. 
That being said, QCD NNLO and EW NLO corrections in di-lepton final states are not expected to change the conclusions of this study. 
New results pertaining to the production of two opposite sign different flavour charged leptons with a full jet veto further confirm the presence of anomalies in similar corners of the leptonic phase-space.}

\begin{document} 

\hfill ICPP-007

\maketitle
\flushbottom

\section{Introduction}
\label{sec:intro}

The incompleteness of the Standard Model (SM) has led the particle physics community to search for a plethora of physics models beyond the SM (BSM).
Owing to a recent study on multi-lepton final states in proton-proton collisions at the LHC~\cite{vonBuddenbrock:2017gvy}, it has become apparent that several anomalous features of the LHC data can be explained through the addition of new scalar bosons to the SM.
In this article we argue that
%, in spite of no discoveries being claimed by the ATLAS and CMS collaborations at the LHC, 
there may already be significant signs of discrepancy in the available LHC data in the area of multiple lepton production.

An early study in 2015 considered the possibility of a heavy scalar $H$ being compatible with several anomalous LHC Run 1 measurements~\cite{vonBuddenbrock:2015ema}.
The result of this study had shown that with a single parameter \bgs (the scale factor for the production cross section of $H$) a set of ATLAS and CMS physics results could be fit with a significance of $3\sigma$.
The square root of the fit parameter was constrained to a value of $\bg=1.5\pm0.6$, which pertained dominantly to the production of the SM Higgs boson ($h$) in association with two dark matter particles $\chi$ from the decay process $H\to h\chi\chi$ through an effective vertex. The best fit mass of $H$ was found to be at $m_H=272^{+12}_{-9}$~GeV. This study included the production of multiple leptons in association with $b$-jets, as reported by the search for the SM Higgs boson in association with top quarks. Other multi-lepton final states considered in Ref.~\cite{vonBuddenbrock:2017gvy} and here were not included in the significance reported in Ref.~\cite{vonBuddenbrock:2015ema}. Now it seems evident that some of the multi-lepton final states considered in  Ref.~\cite{vonBuddenbrock:2017gvy} and here displayed sings of discrepancies with respect to SM predictions already in Run 1 data sets.\footnote{Early discrepancies from Run 1 data sets not considered in  Ref.~\cite{vonBuddenbrock:2015ema} include opposite sign di-leptons and missing transverse energy with a full hadronic jet veto (see Ref.~\cite{vonBuddenbrock:2017gvy}) or di-leptons in association with at least one $b$-jet~\cite{Aad:2014mfk,Khachatryan:2016xws}, among others (see also~\Cref{sec:atlas_exot_2013_16} and~\Cref{sec:atlas_topq_2015_02}). }

Following a discussion of the results in Ref.~\cite{vonBuddenbrock:2015ema}, the next point of interest was to explore the possibility of introducing a scalar mediator $S$ (instead of using effective vertices), such that $H$ could decay to $Sh$, $SS$, and $hh$~\cite{vonBuddenbrock:2016rmr}.
The $S$ was assumed to have globally re-scaled Higgs-like couplings, such that its branching ratios (BRs) could be fixed. In this setup, and in the light of the results in Ref.~\cite{Aaboud:2017uak} where the 100\% branching ratio of $S$ into Dark Matter was ruled out, multi-lepton final states became a focus.
The possibility of embedding $H$ into a Type-II two Higgs doublet model (2HDM) was also discussed, where allowed parameter space of the model was reported in Ref.~\cite{vonBuddenbrock:2018xar}.
More importantly, however, a predictive set of potential search channels for the new scalars was shown.
Several of these predictions were tested and expanded upon in Ref.~\cite{vonBuddenbrock:2017gvy}, where the $\bgs$ parameter was constrained to the value of $1.38\pm 0.22$ through a simultaneous fit to several independent data sets.

In a similar spirit, this article presents the results of an updated fit to the available relevant ATLAS and CMS data with final states containing multiple leptons.
The same simplified model is used along with several assumptions, as described in \Cref{sec:model}, which acts as a source of multiple lepton production in association with $b$-tagged jets ($b$-jets).
The masses of the new scalars are fixed and so only a single degree of freedom is necessary to be constrained (that is, $\bgs$).
Using the simplified model, events are generated and analysed for a statistical comparison with the experimental results.
The specific details of these procedures are outlined in~\Cref{sec:analysis}.
The ensemble of experimental results under consideration in this article is then discussed in \Cref{sec:results}, and thereafter a fit is made to each result with the BSM prediction of the simplified model considered in this article.
A combination fit for the entire ensemble of results is also shown.
Finally in \Cref{sec:discussion}, the successes and failures of the introduced simplified model are discussed in light of the fit results.
Suggestions and prospects for the future of this work are also discussed.

\section{The simplified model\label{sec:model}}

Without appealing to any ultraviolet complete theory,\footnote{It can be shown, however, that the assumptions used to construct the model can be mapped consistently onto the parameter space of a 2HDM with an additional singlet scalar~\cite{vonBuddenbrock:2018xar}.} the anomaly-free simplified model considered in this article is constructed from a few simple assumptions.
Most importantly, we postulate the existence of two new scalar bosons, $H$ and $S$.
The masses of $H$ and $S$ are considered to be fixed in this article, and take on the values of $m_H=270$~GeV and $m_S=150$~GeV.
These choices are based on the best-fit values obtained for the masses from previous studies~\cite{vonBuddenbrock:2015ema,vonBuddenbrock:2017gvy}.
Fixing these values according to the results of completely independent data sets is useful for us to avoid potential bias in tuning the masses to replicate excesses in the data.
Therefore, a ``look elsewhere effect'' is not needed to quantify a global significance in fits, since no other masses are considered.

As alluded to in \Cref{sec:intro}, the mass of $H$ is the result of a primitive study on LHC Run 1 data including distortions in the Higgs boson \pT spectrum~\cite{Aad:2014lwa,Aad:2014tca,Khachatryan:2015rxa,Khachatryan:2015yvw}, di-Higgs and di-boson resonance searches~\cite{Aad:2015xja,Khachatryan:2016sey,Khachatryan:2015tha,Khachatryan:2014jya,Aad:2015agg,Aad:2015kna,Khachatryan:2015cwa}, and measurements on the rate of top associated Higgs production~\cite{Aad:2014lma,Aad:2015iha,Aad:2015gra,Khachatryan:2014qaa}.
The fit procedure is described in Ref.~\cite{vonBuddenbrock:2015ema}, but essentially considers a new associated production mode of $h$ through the decay of $H$.
Seeing as though this was made on an early set of LHC results, it is quite likely that the statistical significance of the fit (which was around $3\sigma$ at a mass of $m_H=270$~GeV) was influenced by fluctuations in the data  -- see for instance the magnitude of the distortions in the ATLAS Higgs boson \pT spectra~\cite{Aad:2014lwa,Aad:2014tca}; more recent independent measurements have shown much milder distortions around lower values of \pT~\cite{ATLAS:2019mju,CMS:2019chr,Sirunyan:2018kta}.
However, this is still consistent with the initial mass result of 270~GeV and in order to be unbiased towards tuning in this article, this value is used.
Similarly, the mass of $S$ is fixed to 150~GeV based on the best-fit point in the statistical study performed in Ref.~\cite{vonBuddenbrock:2017gvy}.
The results considered for the statistical study were all LHC Run 1 measurements of the di-lepton invariant mass (\mll) spectrum in different categories of jet and $b$-tagged jet multiplicity~\cite{Aaboud:2017ujq,Aad:2016wpd,Aaboud:2016mrt,Khachatryan:2015sga,Chatrchyan:2013faa}.
In these results, the mass of $S$ can be weakly constrained by fitting to deviations in the data at low \mll (that is, \mll lower than about 60~GeV).

In this paper, the decays of $H$ and $S$ are not deduced from the data.
Rather, they are fixed to simplified benchmark choices that can be re-interpreted if necessary.
For $H$, we have decided to consider the $H\to Sh$ decay mode as completely dominant, with a 100\% BR.
This is approximately consistent with what was found in Ref.~\cite{vonBuddenbrock:2015ema}, that the BRs of $H$ to SM particles and Higgs boson pairs is small (of the order of 10\% and less) and the dominant BR is to a mode that produces a Higgs boson in associated with additional particles, i.e.~$H\to Sh$.\footnote{In the case of Ref.~\cite{vonBuddenbrock:2015ema}, it was $H\to h\chi\chi$, where the two dark matter particles $\chi$ can be mediated by $S$. This idea is explored in Ref.~\cite{vonBuddenbrock:2016rmr}.}
In terms of the $S$ BRs, since there is no clear choice to make in terms of its decays, it is given Higgs-like BRs through a choice of its couplings.
This is discussed below in \Cref{sec:formalism}, and later in this article it is argued that this simple choice can not fully explain the deviations that are argued to be seen in the data.

\subsection{Formalism}
\label{sec:formalism}

In terms of interactions, $H$ is assumed to be linked to electro-weak symmetry breaking in that it has Yukawa couplings and tree-level couplings the the weak vector bosons $V$ ($W^\pm$ and $Z$).
After electro-weak symmetry breaking, the Lagrangian describing $H$ is Higgs boson-like.
Omitting the terms that are irrelevant in this analysis, $H$ interacts with the SM particles in the following way:
\begin{equation}
\mathcal{L}_{\text{int}} \supset -\beta_{g}\frac{m_t}{v}t\bar{t}H + \beta_{_V}\frac{m_V^2}{v}g_{\mu\nu}~V^{\mu}V^{\nu}H.  \label{eqn:H_production}
\end{equation}
These are the the Higgs-like couplings for $H$ with the top quark ($t$) and the weak vector bosons, respectively.
The strength of each of the couplings is controlled by a free parameter: $\beta_g$ for the $H$-$t$-$t$ interaction and $\beta_V$ for the $H$-$V$-$V$ interaction.
The vacuum expectation value $v$ has a value of approximately $246$~GeV.
The omitted terms include the Yukawa couplings to the other SM fermions and self-interaction terms for $H$.
It can be expected that the couplings to the other SM fermions would also differ by a factor like \bg, however the effect would not make a noticeable difference to the analysis considered in this article and therefore these terms are neglected.
Such numbers could also not be deduced from the LHC data at its current reach, but could be considered with future searches for $H\to b\bar{b}$ and $\mu^+\mu^-$, for example.

\begin{figure}
    \centering
    \begin{subfigure}[b]{\textwidth}
    \centering
    \begin{fmffile}{ggf}
	        \begin{fmfgraph*}(150,120)
	            \fmfstraight
	            \fmfleft{g1i,g2i}
	            \fmfright{p1,ho,p2,so,p3}
	            \fmf{gluon,tension=2}{g1i,g1t}
	            \fmf{gluon,tension=2}{g2i,g2t}
	            \fmf{fermion}{g2t,g1t}
	            \fmfv{label=$\sim\bg$,label.angle=-70}{ttH}
	            \fmf{fermion}{g1t,ttH}
	            \fmf{fermion}{ttH,g2t}
	            \fmfdot{ttH}
	            \fmf{dashes,tension=1.5,label=$H$,l.side=left}{ttH,Hsh}
	            \fmf{dashes,tension=1.5}{Hsh,so}
    	        \fmf{dashes,tension=1.5}{Hsh,ho}
	            \fmflabel{$S$}{so}
	            \fmflabel{$h$}{ho}
	        \end{fmfgraph*}
	    \end{fmffile}
	    \caption{Gluon fusion (\ggf).}
	    \label{fig:ggf}
	\end{subfigure}
	
	\begin{subfigure}[b]{0.45\textwidth}
    \centering
    \begin{fmffile}{tth}
	        \begin{fmfgraph*}(120,150)
	            \fmfstraight
	            \fmfleft{p1,g1i,pm,g2i,p2}
	            \fmfright{pr1,t1o,ho,so,t2o,pr2}
	            \fmf{gluon,tension=1.5}{g1i,g1t}
	            \fmf{gluon,tension=1.5}{g2t,g2i}
	            \fmf{fermion}{t1o,g1t}
	            \fmfv{label=$\sim\bg$,label.angle=180}{ttH}
	            \fmf{fermion}{g1t,ttH}
	            \fmf{fermion}{ttH,g2t}
	            \fmfdot{ttH}
	            \fmf{fermion}{g2t,t2o}
	            \fmf{dashes,label=$H$,l.side=left}{ttH,Hsh}
	            \fmf{dashes}{Hsh,so}
    	        \fmf{dashes}{Hsh,ho}
	            \fmflabel{$S$}{so}
	            \fmflabel{$h$}{ho}
	            \fmflabel{$\bar{t}$}{t1o}
	            \fmflabel{$t$}{t2o}
	        \end{fmfgraph*}
	    \end{fmffile}
	    \caption{Top pair associated production ($ttH$).}
	    \label{fig:ttH}
	\end{subfigure}
	\hfill
	\begin{subfigure}[b]{0.45\textwidth}
    \centering
    \begin{fmffile}{th}
	        \begin{fmfgraph*}(120,150)
	            \fmfstraight
	            \fmfleft{p1,g1i,pm,g2i,p2}
	            \fmfright{pr1,t1o,ho,so,t2o,pr2}
	            \fmf{fermion}{g1i,g1t}
	            \fmf{fermion}{g2i,g2t}
	            \fmfv{label=$\sim\bg$,label.angle=-120}{ttH}
	            \fmf{fermion}{ttH,t1o}
	            \fmf{plain,tension=2.4}{ttH,g1t}
	            \fmfdot{ttH}
	            \fmf{boson,label=$W^\pm$,l.side=left}{g1t,g2t}
	            \fmf{fermion}{g2t,t2o}
	            \fmf{dashes,label=$H$,l.side=left}{ttH,Hsh}
	            \fmf{dashes}{Hsh,so}
    	        \fmf{dashes}{Hsh,ho}
	            \fmflabel{$S$}{so}
	            \fmflabel{$h$}{ho}
	            \fmflabel{$t$}{t1o}
	            \fmflabel{$j$}{g2i}
	            \fmflabel{$b$}{g1i}
	            \fmflabel{$j^\prime$}{t2o}
	        \end{fmfgraph*}
	    \end{fmffile}
	    \caption{Single top associated production ($tH$).}
	    \label{fig:tH}
	\end{subfigure}
    \caption{The representative Feynman diagrams for the leading order production modes of $H$ and its subsequent decay to $Sh$.
    For the sake of clarity, the $H$-$t$-$t$ vertices have been given a label in order to show how the parameter \bg affects the diagrams.}
    \label{fig:feynman_diagrams}
\end{figure}
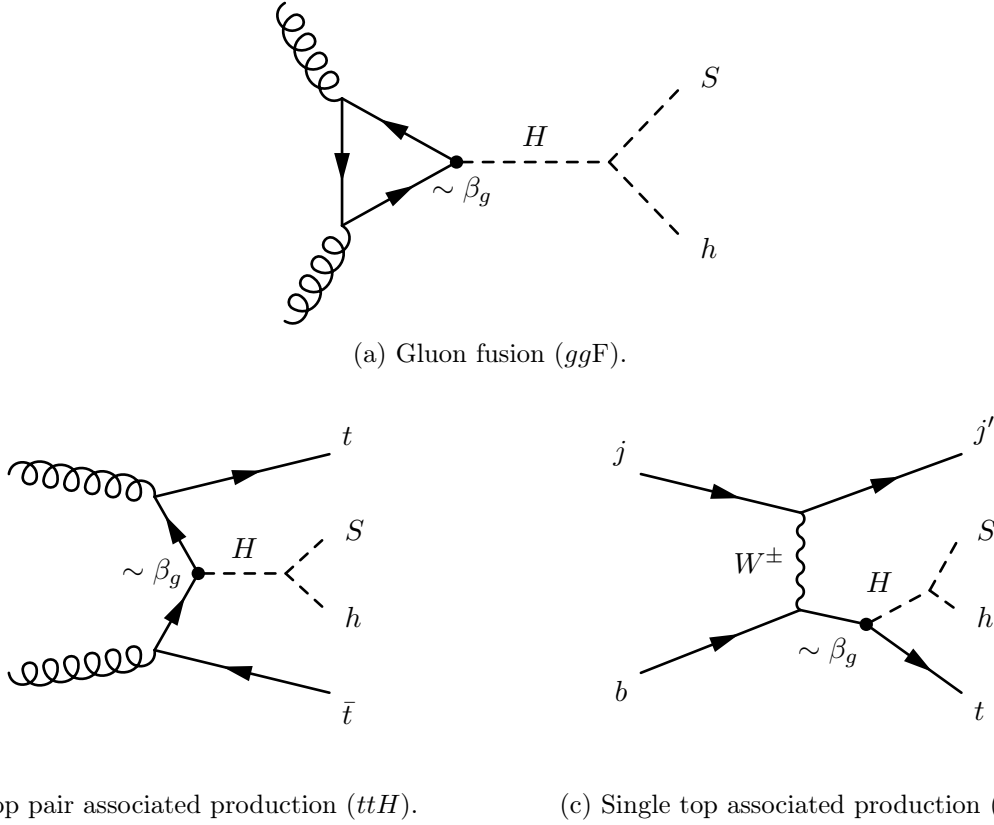

The first term in \Cref{eqn:H_production} allows for the gluon fusion (\ggf) production mode of $H$.
As a baseline, \bg is set to unity such that the $H$ is produced with a Higgs-like cross section.
Due to the squaring of the matrix element in width calculations, production cross sections involving this Yukawa coupling are scaled by $\bgs$.
Therefore, the value of \bgs is used as a free parameter in fits to the data.
We have set $\beta_{_V}=0$, such that the coupling of $H$ to pairs of the weak vector bosons is significantly small; the associated production of $H$ with the weak vector bosons and vector boson fusion (VBF) are negligible production modes.\footnote{A study on the implications of including the VBF production mode is currently underway~\cite{VBF_production}.}
The dominant production mode of $H$ is therefore \ggf, while both single ($tH$) and double ($ttH$) top associated production of $H$ are also non-negligible.
While single top associated production of a Higgs-like boson is usually suppressed due to interference, the implicit assumption of a significantly small $H$-$V$-$V$ coupling allows for a sizeable $tH$ production cross section~\cite{Farina:2012xp}.
It has been shown in previous studies~\cite{vonBuddenbrock:2017gvy,vonBuddenbrock:2015ema} that the $tH$ cross section is enhanced to being approximately that of the $ttH$ cross section.
The representative Feynman diagrams for the production modes of $H$ are shown in \Cref{fig:feynman_diagrams}, along with an indication of how the parameter \bg affects diagrams.

The $S$ boson, on the other hand, is assumed not to be produced directly but rather through the decay of $H$.
In principle, it is possible to include $S$ as a singlet scalar that has interactions with $H$ and the SM Higgs boson $h$.
Doing this would allow the $H$ to produce $S$ bosons through the $H\to SS$ and $Sh$ decay modes.
Here we assume the $H\to Sh$ decay mode to have a 100\% BR (also shown in \Cref{fig:feynman_diagrams}).
These assumptions are all achieved by introducing the following effective interaction Lagrangians.
Firstly, $S$ is given a vacuum expectation value and couples to the scalar sector,
\begin{align}
{\cal L}_{HhS} = &-\frac{1}{2}~v\Big[\lambda_{_{hhS}} hhS + \lambda_{_{hSS}} hSS +
\lambda_{_{HHS}} HHS \notag \\ & + \lambda_{_{HSS}} HSS + \lambda_{_{HhS}} HhS\Big], \label{eqn:HS_coupling}
\end{align}
where the couplings are fixed to ensure that the $H\to Sh$ BR is 100\%.
Secondly, $S$ is given Higgs-like BRs by fixing the parameters in the Lagrangian,
\begin{align}
{\cal L}_{S} =&~ \frac{1}{4} \kappa_{_{Sgg}} \frac{\alpha_{s}}{12 \pi v} S G^{a\mu\nu}G_{\mu\nu}^a
+ \frac{1}{4} \kappa_{_{S\gamma\gamma}} \frac{\alpha}{\pi v} S F^{\mu\nu}F_{\mu\nu} \notag \\
& + \frac{1}{4} \kappa_{_{SZZ}} \frac{\alpha}{\pi v} S Z^{\mu\nu}Z_{\mu\nu}
  + \frac{1}{4} \kappa_{_{SZ\gamma}} \frac{\alpha}{\pi v} S Z^{\mu\nu}F_{\mu\nu} \notag \\
& + \frac{1}{4} \kappa_{_{SWW}} \frac{2 \alpha}{\pi s_w^2 v} S W^{+\mu\nu}W^{-}_{\mu\nu} - \sum_f \kappa_{_{Sf}} \frac{m_f}{v} S \bar f f. \label{eqn:S_decays}
\end{align}

The couplings in \Cref{eqn:S_decays} are chosen to be globally re-scaled Higgs-like couplings.
This is somewhat an arbitrary choice, although it has the dual advantage of fixing the BRs of $S$ (which in turn reduces the number of free parameters in the model) and suppressing the direct production of $S$.
The latter advantage is motivated by the LHC data, since there have been no observations of directly produced Higgs-like bosons near a mass of 150~GeV at the LHC as of yet.
It is possible to determine an upper limit on the value by which the couplings are re-scaled by considering the ATLAS and CMS searches for a Higgs boson in the $h\to ZZ^*\to4\ell$ decay channel~\cite{ATLAS:2019ssu,CMS:2019chr}, which provide event yields as a function of the Higgs boson mass.
By considering the cross section and BR of a Higgs boson at a mass of 150~GeV, assuming that such a particle would have the same reconstruction efficiency and acceptance as the SM Higgs boson, and considering a systematic uncertainty of 10\% on the SM prediction, the LHC data exclude a value larger than $0.23$ at the 95\% confidence level.
Squaring this value gives $0.054$, which is the limit on the relative rate for such a production compared with the un-scaled production rate.
In the analysis done in this paper, the value does not have to be chosen explicitly since only the BRs are needed as inputs, and a direct production mechanism of $S$ would not affect the results.
A few of the shortcomings of making the arbitrary choice for the $S$ couplings are discussed in \Cref{sec:discussion}.

\section{Analysis strategy\label{sec:analysis}}
\subsection{Theoretical predictions\label{sec:predictions}}

Predictions of the shapes produced by the BSM processes described above were constructed from Monte Carlo (MC) events as calculated by different event generators.
For \ggf, the built in matrix elements for BSM Higgs boson production from \texttt{Pythia 8}~\cite{Sjostrand:2014zea} were utilised for the hard scatter process.
For the case of $tH$ and $ttH$, the hard scatter process was simulated at leading order (LO) using \texttt{aMC@NLO}~\cite{Alwall:2014hca}.
In the case of $tH$, a five-flavour proton is used to construct the matrix element.
The resonance decays, parton shower and hadronisation are all performed by \texttt{Pythia 8} for each BSM process.
The masses of the relevant scalars were fixed to $m_H=270$~GeV, $m_S=150$~GeV and $m_h=125.09$~GeV.
The cross sections for each BSM process were scaled to the highest order SM Higgs-like production cross sections at a mass of 270~GeV, as taken from the LHC Higgs cross section working group~\cite{deFlorian:2227475}. 
Such a cross section corresponds to $\bgs=1$.
Following the discussion in \Cref{sec:model}, the $tH$ cross section is set to be the same as the given $ttH$ cross section.

The generated events are then passed through the \texttt{Delphes 3}~\cite{deFavereau:2013fsa} fast simulation package to model the response of the appropriate detector.
This is done in conjunction with an event selection using the \texttt{CheckMATE 2}~\cite{Dercks:2016npn} analysis framework (which, by design, uses the \texttt{FastJet}~\cite{Cacciari:2011ma} method to reconstruct jets).
Event selections were designed by hand to replicate the cuts and analysis techniques given in the experimental search results.
The selection codes were validated by ensuring that the selection for a chosen process given in the experimental publications could be reproduced to within 10\% of the value obtained by the custom designed selection codes used for the results in this article.
Given the known discrepancies between the \texttt{Delphes 3} fast simulation and the full simulation done by the proprietary ATLAS and CMS software, the resulting distributions from the validation process performed very well when compared to the given distributions as shown by the experimental collaborations.

In order to maintain consistency with the experimental results, the SM background predictions and their associated systematic uncertainties were taken directly from the experimental publications.
In several cases, additional systematic uncertainties were applied to the SM background predictions; these shall be explained in detail for each individual analysis in \Cref{sec:results}.

\subsection{Statistical tools\label{sec:stats}}

All fits to the LHC data in this article were performed using the \texttt{HistFactory} extension of the \texttt{RooStats} framework~\cite{Cranmer:1456844}, which is a template-based method of performing fits based on maximising a profile likelihood ratio.
The SM components of the fits are always taken directly from the published experimental distributions, along with their associated systematic uncertainties, which can be incorporated into the \texttt{HistFactory} schema as variations of a distribution's normalisation or shape.
The BSM component is always constructed using a single mass point ($m_H=270$~GeV and $m_S=150$~GeV) and therefore has only one degree of freedom under the assumptions stated in \Cref{sec:model}.
The single degree of freedom is $\bgs$, which maps directly to the normalisation of the BSM signal with respect to the SM Higgs-like production cross section of $H$.

The statistical likelihood function $L\left(\bgs~|~\theta\right)$ is constructed as the product of Poisson probabilities for each bin and in each considered measurement.
Systematic uncertainties are incorporated as additional constraint factors in the likelihood, which vary according to their associated nuisance parameters $\theta$ (Ref.~\cite{Cranmer:1456844} contains a full description of how the likelihood is parameterised in terms of different kinds of systematic uncertainties).
The general form of the profile likelihood ratio then takes the form,
\begin{equation}
    \lambda\left(\bgs\right)=\frac{L\left(\bgs~|~\doublehat{\theta}\right)}{L\left(\hat{\beta}_g^2~|~\hat{\theta}\right)},
    \label{eqn:plr}
\end{equation}
where $\doublehat{\theta}$ is the set of nuisance parameters which maximise the likelihood function for a given value of $\bgs$, and $\hat{\beta}_g^2$ and $\hat{\theta}$ are the values of $\bgs$ and the set of nuisance parameters which maximise the likelihood function over the entire parameter space.
The best-fit value of the parameter of interest $\bgs$ is usually identified as the minimum of $-2\log\lambda\left(\bgs\right)$, where a deviation of one unit in this quantity is equivalent to a $1\sigma$ deviation from the best-fit point of the parameter of interest.
Since the value $\bgs=0$ corresponds to the SM-only hypothesis (the \textit{null} hypothesis), the significance of each fit is calculated as,
\begin{equation}
    Z=\sqrt{-2\log\lambda\left(0\right)}.
    \label{eqn:significance}
\end{equation}

\section{Fits to LHC data\label{sec:results}}

Experimental searches for final states containing multiple leptons in high energy proton-proton collisions at the LHC have been performed in a variety of contexts by the ATLAS and CMS experiments.
These include searches for the SM production of top quarks decaying to opposite-sign (OS) lepton pairs, searches for Higgs boson production in leptonic final states and BSM searches for the production of same-sign (SS) lepton pairs, to name a few.
Many of these searches involve either a signal or dominant background component that contains top quarks in the final state.
Therefore, the results are often always dependent on the number of $b$-jets produced with the leptons.
Note that in this article any lepton $\ell$ refers to either an electron or a muon. 
Contributions from the production of $\tau$-leptons are only relevant if they subsequently decay leptonically.

\begin{table}
    \centering
    \begin{tabular}{l|l|l}
    \hline
        \multicolumn{1}{c|}{\textbf{Data set}} & \multicolumn{1}{|c|}{\textbf{Reference}} & \multicolumn{1}{|c}{\textbf{Selection}}  \\
        \hline
        ATLAS Run 1 & ATLAS-EXOT-2013-16\hfill\cite{Aad:2015gdg} & SS $\ell\ell$ and $\ell\ell\ell$ + $b$-jets \\
        ATLAS Run 1 & ATLAS-TOPQ-2015-02\hfill\cite{Aaboud:2017ujq} & OS $e\mu$ + $b$-jets \\
        CMS Run 2 & CMS-PAS-HIG-17-005\hfill\cite{CMS-PAS-HIG-17-005} & SS $e\mu$, $\mu\mu$ and $\ell\ell\ell$ + $b$-jets \\
        CMS Run 2 & CMS-TOP-17-018\hfill\cite{Sirunyan:2018lcp} & OS $e\mu$ \\
        CMS Run 2 & CMS-PAS-SMP-18-002\hfill\cite{CMS-PAS-SMP-18-002} & $\ell\ell\ell+\MET$ ($WZ$) \\
        ATLAS Run 2 & ATLAS-EXOT-2016-16\hfill\cite{Aaboud:2018xpj} & SS $\ell\ell$ and $\ell\ell\ell$ + $b$-jets \\
        ATLAS Run 2 & ATLAS-CONF-2018-027\hfill\cite{ATLAS-CONF-2018-027} & OS $e\mu$ + $b$-jets \\
        ATLAS Run 2 & ATLAS-CONF-2018-034~\hfill\cite{ATLAS-CONF-2018-034} & $\ell\ell\ell+\MET$ ($WZ$) \\
        \hline
    \end{tabular}
    \caption{A list of the ATLAS and CMS experimental results pertaining to final states with multiple leptons that are considered in this article.
    For each result, a simple baseline selection is shown.
    The different kinematic cuts and categories are not shown here, but are described for each analysis below.}
    \label{tab:results_list}
\end{table}

The ensemble of results considered in this article is shown in \Cref{tab:results_list}.
The majority of results come from the Run 2 data sets, due to the fact that the increased luminosity and cross sections of most of the processes implies a greater statistical precision in the data.
The selection of charges for the leptons ensures that each data set is statistically independent, where any potential double counting could only arise through charge mis-identification, and is expected to be negligible.
For each result in \Cref{tab:results_list}, a fit is made using the SM and BSM theoretical predictions discussed in \Cref{sec:predictions} as inputs to the statistical method described in \Cref{sec:stats}.
The results of each fit are shown in the sections below.
In \Cref{sec:combination}, a combination of all the results is shown.

\subsection{ATLAS Run 1 search for SS leptons in association with \texorpdfstring{$b$}{b}-jets\label{sec:atlas_exot_2013_16}}

The production of two SS leptons is a rare process in the SM.
This makes it a striking signature for BSM theories that could predict SS lepton pairs via cascaded decays.
The ATLAS Run 1 data set was used in a search for SS lepton pairs in association with $b$-jets, with the goal of constraining BSM models that predict heavy vector-like quarks (VLQs)~\cite{Aad:2015gdg}.
This kind of search is sensitive to the $ttH$ and $tH$ production modes of the simplified model considered in this article, since a SS lepton pair can be selected from the combination leptonic top quark decays and $S\to VV$ decays.
The $b$-jets from the top quark and $h\to bb$ decays make for a high probability of reconstructing three $b$-jets in the final state.

The data set for the search is statistically limited, and therefore the overall rates per signal region (SR) are used to fit the BSM prediction in this case (instead of the differential distributions).
As a baseline selection, the analysis requires two or three leptons in the final state, with at least one SS lepton pair.
The SRs are separated by $b$-jet multiplicity and different cuts on missing transverse energy ($\MET$) and $\HT$, which is the scalar sum of the lepton and jet transverse momenta.
These cuts are optimised to identify the signal from a model that predicts the production of VLQs, but are still sensitive to the simplified model used in this article due to relatively low cuts on $\MET$.
The SRs are defined as follows,
\begin{itemize}[leftmargin=1.2in,noitemsep,topsep=0pt,parsep=0pt,partopsep=0pt]
    \item [SRVLQ0:] \makebox[2.4cm][l]{$\Nb=1$;}\makebox[4.3cm][l]{$\MET>40$~GeV;}$400<\HT<700$~GeV,
    \item [SRVLQ1:] \makebox[2.4cm][l]{$\Nb=2$;}\makebox[4.3cm][l]{$\MET>40$~GeV;}$400<\HT<700$~GeV,
    \item [SRVLQ2:] \makebox[2.4cm][l]{$\Nb\geq3$;}\makebox[4.3cm][l]{$\MET>40$~GeV;}$400<\HT<700$~GeV,
    \item [SRVLQ3:] \makebox[2.4cm][l]{$\Nb=1$;}\makebox[4.3cm][l]{$40<\MET<100$~GeV;}$\HT\geq700$~GeV,
    \item [SRVLQ4:] \makebox[2.4cm][l]{$\Nb=1$;}\makebox[4.3cm][l]{$\MET\geq100$~GeV;}$\HT\geq700$~GeV,
    \item [SRVLQ5:] \makebox[2.4cm][l]{$\Nb=2$;}\makebox[4.3cm][l]{$40<\MET<100$~GeV;}$\HT\geq700$~GeV,
    \item [SRVLQ6:] \makebox[2.4cm][l]{$\Nb=2$;}\makebox[4.3cm][l]{$\MET\geq100$~GeV;}$\HT\geq700$~GeV,
    \item [SRVLQ7:] \makebox[2.4cm][l]{$\Nb\geq3$;}\makebox[4.3cm][l]{$\MET>40$~GeV;}$\HT\geq700$~GeV.
\end{itemize}

\begin{figure}
    \centering
    \includegraphics[width=0.7\textwidth]{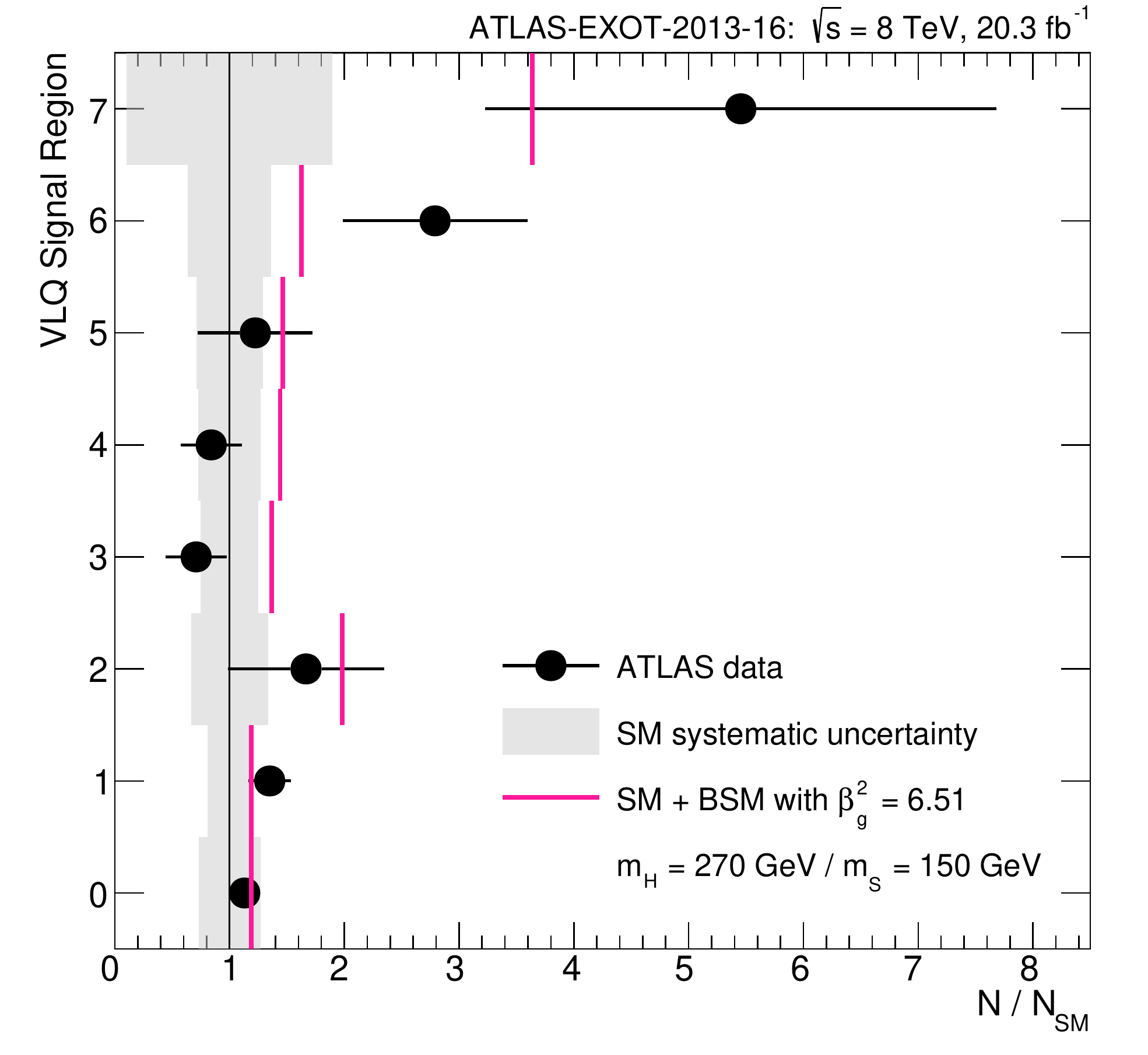}
    \caption{The SM+BSM fit result for the ATLAS Run 1 search for SS leptons in association with $b$-jets binned by the SRs defined in the text.
    The horizontal axis represents a production rate normalised to that of the SM.
    The BSM prediction is scaled to the best-fit value of $\beta_g^2$, while the SM and its systematic uncertainty do not incorporate the constrains imposed by the fitting procedure.}
    \label{fig:atlas_exot_2013_16_plot}
\end{figure}

Given the categorisation shown above, a fit was made to the data using the SM+BSM hypothesis on the total rate per SR.
An overall normalisation systematic uncertainty was applied to the SM prediction, ranging between 19\% and 90\% depending on the SR in question.
This systematic incorporates all the relevant uncertainties on the SM background, including theoretical uncertainties related to the event generators used, as well as experimental uncertainties.
The systematic uncertainty was assumed to be correlated over all of the SRs (i.e. the fit did not allow a bin-by-bin variation of the SM prediction).
In the fit, the BSM fit parameter \bgs was best fit at a value of $6.51\pm2.99$.
This result is relatively high compared to the other results calculated in this article.
The implications of this are discussed in \Cref{sec:discussion}.
In terms of the significance of this deviation from the SM-only hypothesis, this corresponds to $Z=2.37\sigma$.
The performance of the fit per SR can be seen in \Cref{fig:atlas_exot_2013_16_plot}.

Due to a lack of statistics and large systematic uncertainties in this measurement, the significance of the fit is not high enough for this measurement to be noteworthy on its own.
However, the kinematic requirements for each SR shows the simplified model's strength in being able to describe excesses in the data with multiple leptons and at least three $b$-jets.
The requirements on \MET and \HT are relatively loose in this measurement, compared to other experimental searches that consider the production of heavy particles -- super-symmetry (SUSY) for example.
The simplified model in this article is produced dominantly in the region of  the phase space with low \MET, making these searches of particular interest, as opposed to SUSY searches where the model does not produce a significant signal.
The CMS and ATLAS Run 2 versions of this measurement are discussed in \Cref{sec:cms_pas_hig_17_005} and \Cref{sec:atlas_exot_2016_16}, respectively.

\subsection{ATLAS Run 1 di-lepton invariant mass spectrum\label{sec:atlas_topq_2015_02}}

Measurements related to the SM production of top quarks are not typically considered in the search for BSM physics.
However, the very simple selection applied to the events considered in such measurements makes for a set of robust distributions against which new physics theories can be tested.
From the ATLAS Run 1 data set, a set of differential distributions pertaining to the SM production of top quarks was reported~\cite{Aaboud:2017ujq}.
The events selection in this measurement is a simple selection of an electron and a muon in association with at least one $b$-jet.

For this measurement, we have considered \mll as a discriminating variable in the fit.
This is because the simplified model used in this article produces a well defined narrow peak at a value of $\mll\simeq50$~GeV, and can be easily distinguished from the SM background peak at around 90~GeV.
In addition to this, it is modelled relatively consistently with different event generators in the region where the BSM signal is concentrated.
The slight variations of the SM prediction in this region can be covered easily with a systematic uncertainty that affects the SM normalisation.

A fit to this distribution was previously made using the same simplified model in Ref.~\cite{vonBuddenbrock:2017gvy}.
However, the previous fit was made only by maximising a likelihood for the entire distribution; the effects of systematic uncertainties were inferred after the fit result.
Here we present a fit that takes systematic uncertainties into account by minimising the profile likelihood ratio in \Cref{eqn:plr}.
The applied systematic uncertainty is determined as follows.
With the assumption that no significant new physics signals appear in the tail of the distribution, the entire SM prediction is scaled to the data in the region where $\mll>110$~GeV; the scale factor was calculated to be 0.984.
In doing this, many of the systematic uncertainties that affect the normalisation of the SM prediction become irrelevant.
The uncertainties which are not affected by the scaling were then added up in quadrature and found to affect the normalisation of the SM by just under 2\%.
Therefore, a normalisation systematic uncertainty of 2\%, correlated over all of the bins of the distribution is applied to the SM prediction in the fit.
In order to avoid bias with the scaling procedure, the fit is only performed on the bins where $\mll\leq110$~GeV.
The 2\% uncertainty on the normalisation includes the variation of scales for the SM prediction, which only varies by a normalisation factor in the fit region.
More details on this scale uncertainty and the other factors which affect the SM $t\bar{t}$ prediction are shown in \Cref{app:top}.

\begin{figure}
    \centering
    \includegraphics[width=0.88\textwidth]{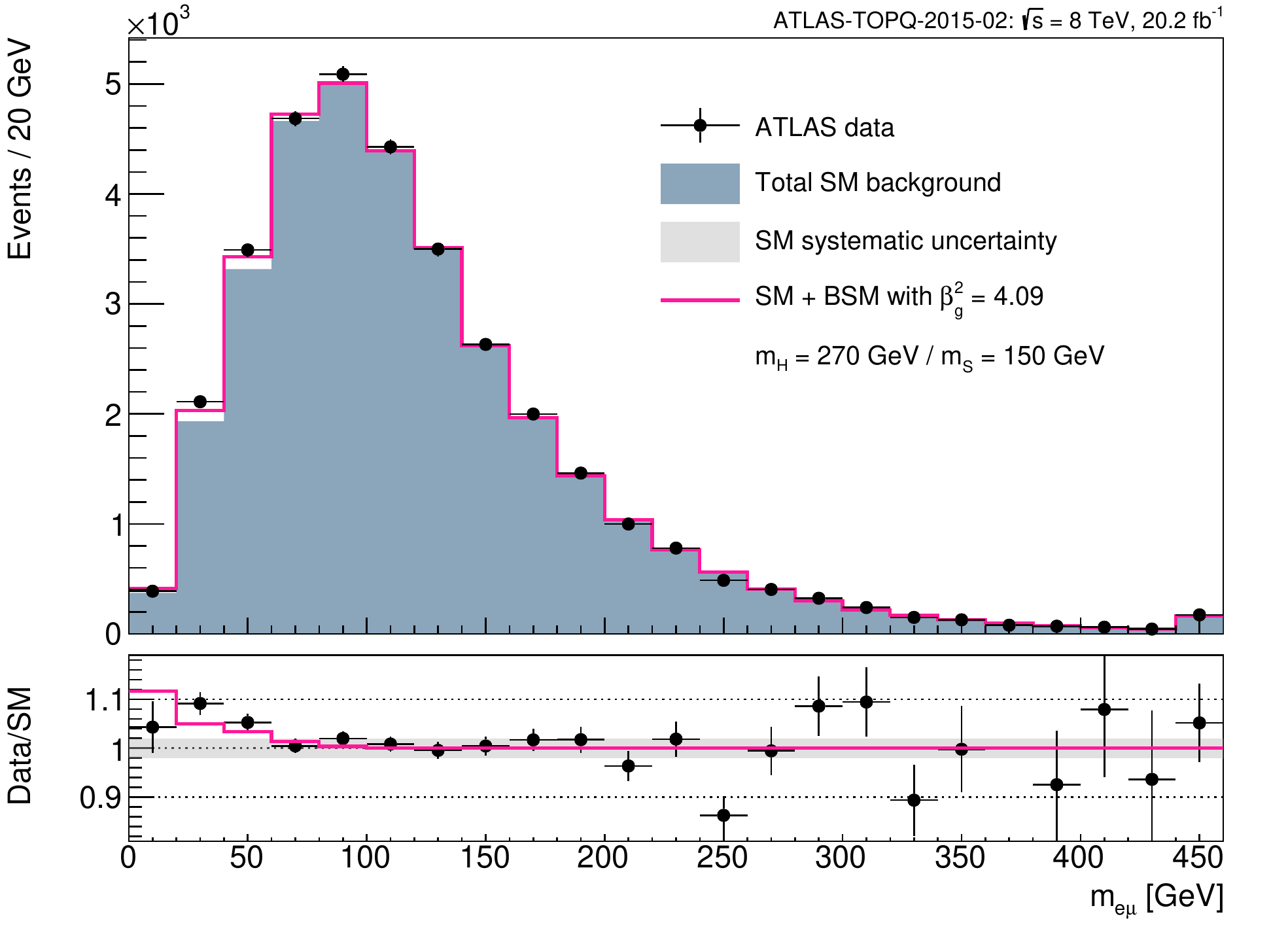}
    
    \includegraphics[width=0.88\textwidth]{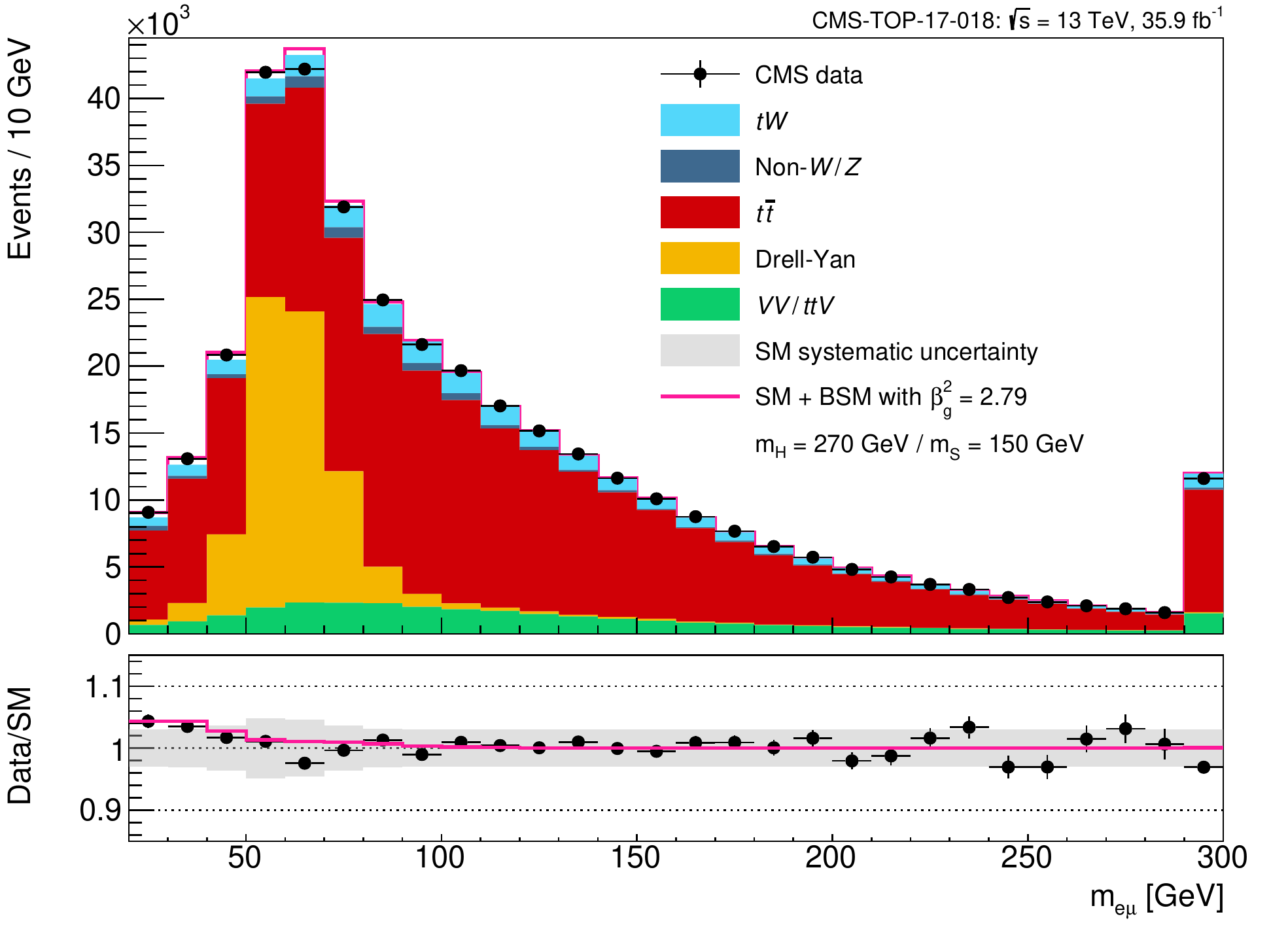}
    \caption{The SM+BSM fit result for the di-lepton invariant mass spectrum reported by ATLAS in Run 1 (above) and CMS in Run 2 (below).
    In each case, \bgs has been scaled to its best-fit value.
    The SM predictions and systematic uncertainties do not incorporate the constraints imposed by the fit.}
    \label{fig:mll_plots}
\end{figure}

The result of the fit gives a best-fit value of \bgs at $4.09\pm1.37$, corresponding to a significance of $Z=2.99\sigma$.
The distribution overlaid with the SM+BSM fit is shown in the upper panel of \Cref{fig:mll_plots}.
The BSM prediction performs very well in its ability to explain the excess in the first few bins of the distribution.
In the fit, the SM prediction is raised by slightly less than 1\% in the profiling of the systematic uncertainty such that the peak of the SM distribution is also fit well.
It should be noted that the tail of the SM distribution, while not included in the fit, is still compatible with this constraint.

\subsection{CMS Run 2 opposite sign di-lepton invariant mass spectrum\label{sec:cms_top_17_018}}

Similar to the ATLAS Run 1 di-lepton invariant mass spectrum discussed in \Cref{sec:atlas_topq_2015_02}, CMS have more recently published an \mll spectrum with a partial Run 2 data set~\cite{Sirunyan:2018lcp}.
The key difference between the ATLAS Run 1 result and the CMS Run 2 result is that CMS does not place any requirements on the number of $b$-jets in the final state.
The CMS result therefore has a significant contribution from the SM Drell-Yan process, which enhances the SM background and shifts the inclusive peak to around 70~GeV.
The fitting procedure for this measurement is similar to that of the ATLAS Run 1 results described in \Cref{sec:atlas_topq_2015_02}, however with a few caveats that shall be discussed below.

On the inspection of the \mll distribution in Ref.~\cite{Sirunyan:2018lcp}, it is clear that there exists an excess of events at low invariant mass values, consistent with the results discussed in \Cref{sec:atlas_topq_2015_02}.
However, it also becomes evident that the entire background SM prediction is poorly modelled with respect to the data, the tail of the distribution is underestimated by the theoretical prediction.
We believe that this is due to a discrepancy present in the nominal $t\bar{t}$ MC prediction used in the measurement (that is, the \texttt{POWHEG V2}~\cite{Alioli:2010xd} $t\bar{t}$ sample).
This is justified by noting the \mll differential distribution in the measurement of $t\bar{t}$ fiducial cross sections~\cite{CMS-PAS-TOP-17-014}, which corresponds to a fiducial phase space enriched by $t\bar{t}$ events.
One can note from this distribution that the same \texttt{POWHEG V2} sample does not describe the data as well as the other two $t\bar{t}$ samples it is tested against.
For this reason, we have used the distribution in Ref.~\cite{CMS-PAS-TOP-17-014} to re-weight the $t\bar{t}$ distribution in Ref.~\cite{Sirunyan:2018lcp}.
In doing so, the large excess at low values of \mll is reduced, and the tail is flattened such that the entire distribution is able to describe the data far better after the re-weighting (up to a normalisation factor).
The origin of this apparent discrepancy in the modelling of the $t\bar{t}$ process is uncertain, and does not seem to be a problem in any of the ATLAS $t\bar{t}$ samples.

The theoretical SM MC predictions are scaled to the data in the region where $\mll>110$~GeV, and the fit is done in the region where $\mll<110$~GeV (similarly to the ATLAS Run 1 \mll distribution).
The Drell-Yan prediction, however, is not altered by this scaling.
Instead, the Drell-Yan prediction is left at its nominal normalisation, and an exceptionally large normalisation systematic uncertainty of 6.86\% is applied to it.
All of the other SM components are given a normalisation systematic uncertainty of 3\%.
Similarly to \Cref{sec:atlas_topq_2015_02}, the scale uncertainties on the $t\bar{t}$ background only affect the normalisation of the distribution in the region where $\mll<110$~GeV, and so the 3\% uncertainty incorporates the scale uncertainties (as described in \Cref{app:top}).
Note also that uncertainties related to the choice of event generator are small as in \Cref{sec:atlas_topq_2015_02}, and are therefore covered by the conservative normalisation uncertainty of 3\%.
The result of the re-weighting and scaling procedures is an \mll distribution that is very well described at the peak and in the tail, however still with a significant excess of events at low \mll.
This can be seen in the lower panel of \Cref{fig:mll_plots}, along with the SM+BSM fit result.

The fitting process favours the SM+BSM hypothesis with a  significance of $Z=5.45\sigma$ (this fit has the highest significance for all of the individual fit results in the ensemble in \Cref{tab:results_list}).
The corresponding best-fit value of $\beta_g^2$ is $2.79\pm0.52$.
The statistical precision of the measurement is the main reason why the significance of the SM+BSM fit is so high.
For one, the statistical uncertainty on the data is negligible compared with the systematic uncertainty on the SM prediction.
In addition to this, the profiling of the systematic uncertainties places strong constraints on the best-fit normalisation of the SM background.
Due to the statistical precision of the data set and the excess of events below $\mll=60$~GeV, there is a strong tension between the SM-only hypothesis and the data.
Since the BSM prediction is distributed exactly where the excess of events is, the SM+BSM fit resolves this tension with a large significance.
It should be noted that any significant variation of the normalisation or shape of the SM background in the fit would have negative consequences on the compatibility of the tail of the distribution with the data, which is so well described by the SM prediction.
In the SM+BSM fit, the variation of the normalisation of the SM backgrounds is negligible (less than 0.1\%).

\begin{figure}
    \centering
    \includegraphics[width=0.88\textwidth]{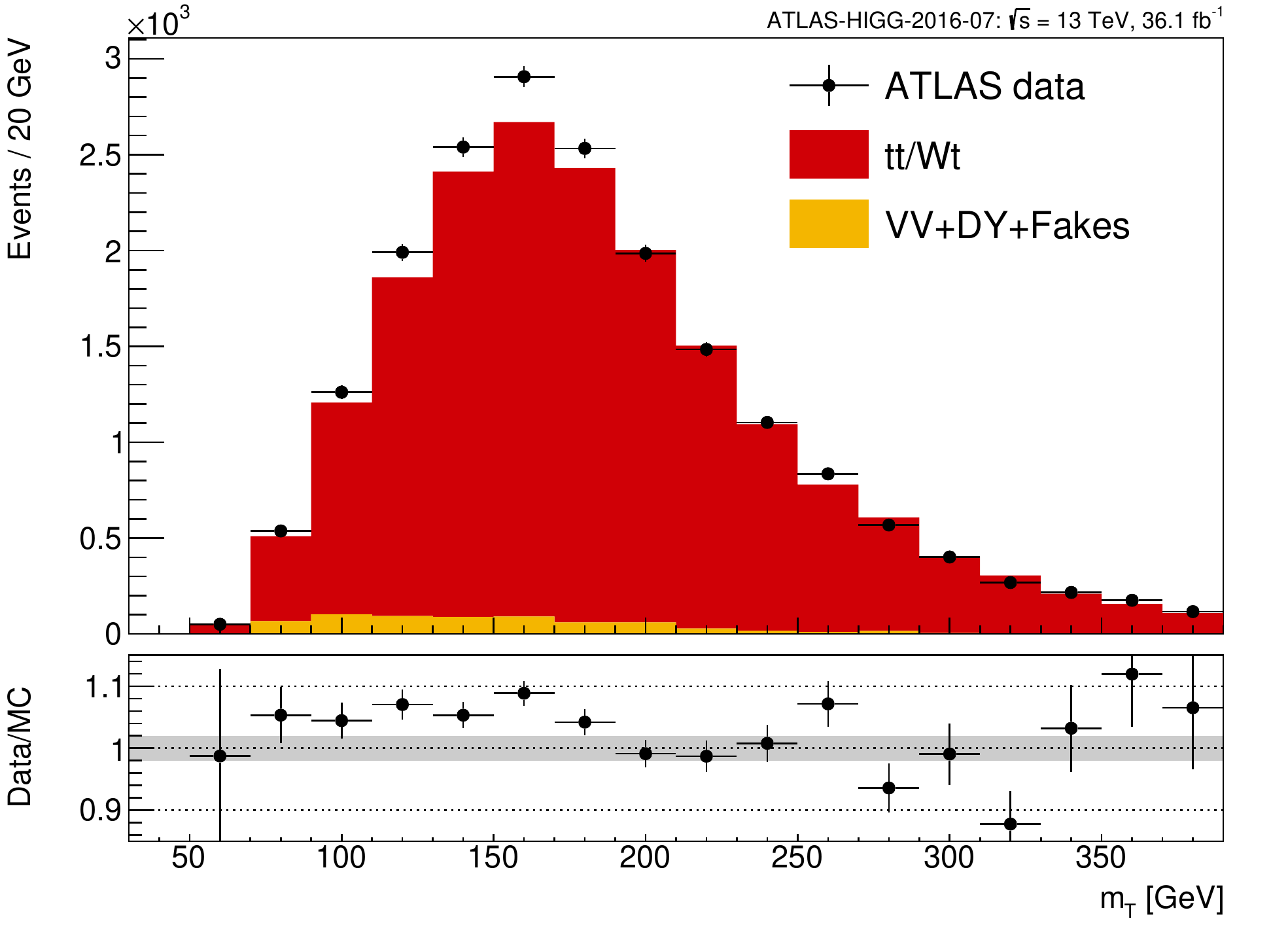}
    \caption{The transverse mass distribution in the ATLAS Run 2 $h\to WW^*$ top CR, scaled such that the integral in data and the SM MC match in the region of $m_\text{T}>200$~GeV.
    An estimated flat systematic uncertainty of 2\% is drawn only as a baseline.}
    \label{fig:atlas_mT}
\end{figure}

With such a significant effect, it was deemed necessary to search for other measurements in which the excess might be localised.
This was found in the $t\bar{t}/Wt$ control region (CR) of the ATLAS Run 2 Higgs production cross section measurement in the $WW^*\to e\nu\mu\nu$ decay channel~\cite{Aaboud:2018jqu}.
The BSM model is sensitive to the selection criteria of this measurement since it requires one high \pT $b$-jet and exactly one central un-tagged jet.
Since the discriminating variable of the search in this measurement is that of transverse mass ($m_\text{T}$), the excess may localise to a broad peak in $S\to WW^*$ decays.
It was decided not to include this measurement in the fit, since its event selection overlaps significantly with that of the measurement discussed in \Cref{sec:atlas_conf_2018_027}, and including it in the combination would be double counting.
%In any case, upon looking at the distributions in the above-mentioned CR, it was noted that the transverse mass ($m_\text{T}$) distribution appeared to be poorly modelled by the SM prediction.
In an attempt to reduce the experimental systematic uncertainties on the distribution, it was determined that the BSM model would only produce a signal in the region of $m_\text{T}<200$~GeV, and therefore the distribution was scaled to match the integral of the data in the region above 200~GeV.
Doing this reveals a broad structure in the data, compared with the SM prediction, that peaks at around $m_\text{T}=150$~GeV.
This can be seen in \Cref{fig:atlas_mT}.
Such a structure could be well described by a resonance decaying to a pair of $W$ bosons in association with a $b$-jet; in this case the $S$ boson (having a mass of 150~GeV) is a prime candidate.
The $m_\text{T}$ distribution, however, does not have a peak where the data peaks.
Due to the off-shell nature of the $H\to Sh$ decay, the distribution peaks below the Higgs mass.\footnote{This also implies that small changes in the masses of $H$ and $S$ would not drastically shift the peak. As long as the $H\to Sh$ decay is off-shell, the position of the peak is saturated towards a value below the Higgs mass.}
The SM+BSM fit still improved on the SM-only hypothesis, however the improvement was weak (less than 2$\sigma$).

MC studies involving $t\overline{t}$ and $Wt$ processes are reported in \Cref{app:top}, where the robustness of the transverse mass against $b$-tagged jet multiplicity and QCD variations is demonstrated.

\subsection{CMS Run 2 search for same-sign leptons in association with \texorpdfstring{$b$}{b}-jets\label{sec:cms_pas_hig_17_005}}

The CMS Run 2 data set was used to make a search for the SM Higgs boson in association with a single top quark~\cite{CMS-PAS-HIG-17-005}.
The event selection in this search is very similar to the ATLAS searches for SS leptons in association with $b$-jets (discussed in \Cref{sec:atlas_exot_2013_16} for Run 1 and below in \Cref{sec:atlas_exot_2016_16} for Run 2).
The simple event pre-selection requires two or three leptons in the final state with at least one SS pair.
In addition, at least one $b$-jet is required along with at least one untagged jet.
The events are further categorised into three categories, $e\mu$, $\mu\mu$ and $\ell\ell\ell$, depending on the multiplicity and flavour of the leptons.

A very simple fit was made using this measurement in the results of Ref.~\cite{vonBuddenbrock:2017gvy}, where a total rate of events for $N_\text{jet}\geq3$ with and without the BSM signal was compared for each category.
Here, the fit is extended by combining all three categories for a greater statistical precision.
The variable that is used to fit the BSM prediction is the highest pseudo-rapidity ($\eta$) for high $p_\text{T}$ jets (that is, jets with $p_\text{T}>40$~GeV).
This decision was made based on the fact that the systematic uncertainty on the SM is smaller for this variable than that of the other variables considered in the search (in addition to a reasonable signal to background discrimination).
The applied systematic uncertainty is considered to be an overall normalisation variation, calculated as the sum of the total systematic variation from each category.

The BSM contamination in this signal region is non-trivial.
Whilst one might expect that the dominant contribution to the event selection comes from the $ttH$ production mode, it is actually the \ggf production mode that is dominant.
This is due to the much larger \ggf cross section compared to that of $ttH$ and $tH$, in addition to the fact that leptons from heavy flavour (HF) decays in the \ggf production mode contribute to the signal with a non-negligible probability.
These effects are usually accounted for in the lepton isolation criteria used by the experiment.
However, due to the ambiguity of these criteria described in this particular measurement, a crude estimate was made on the probabilities for both prompt and HF decay leptons to be accepted or rejected in the selection.
This estimate was validated using event yield for known processes given in the article, and was found to perform well.

\begin{figure}
    \centering
    \includegraphics[width=0.49\textwidth]{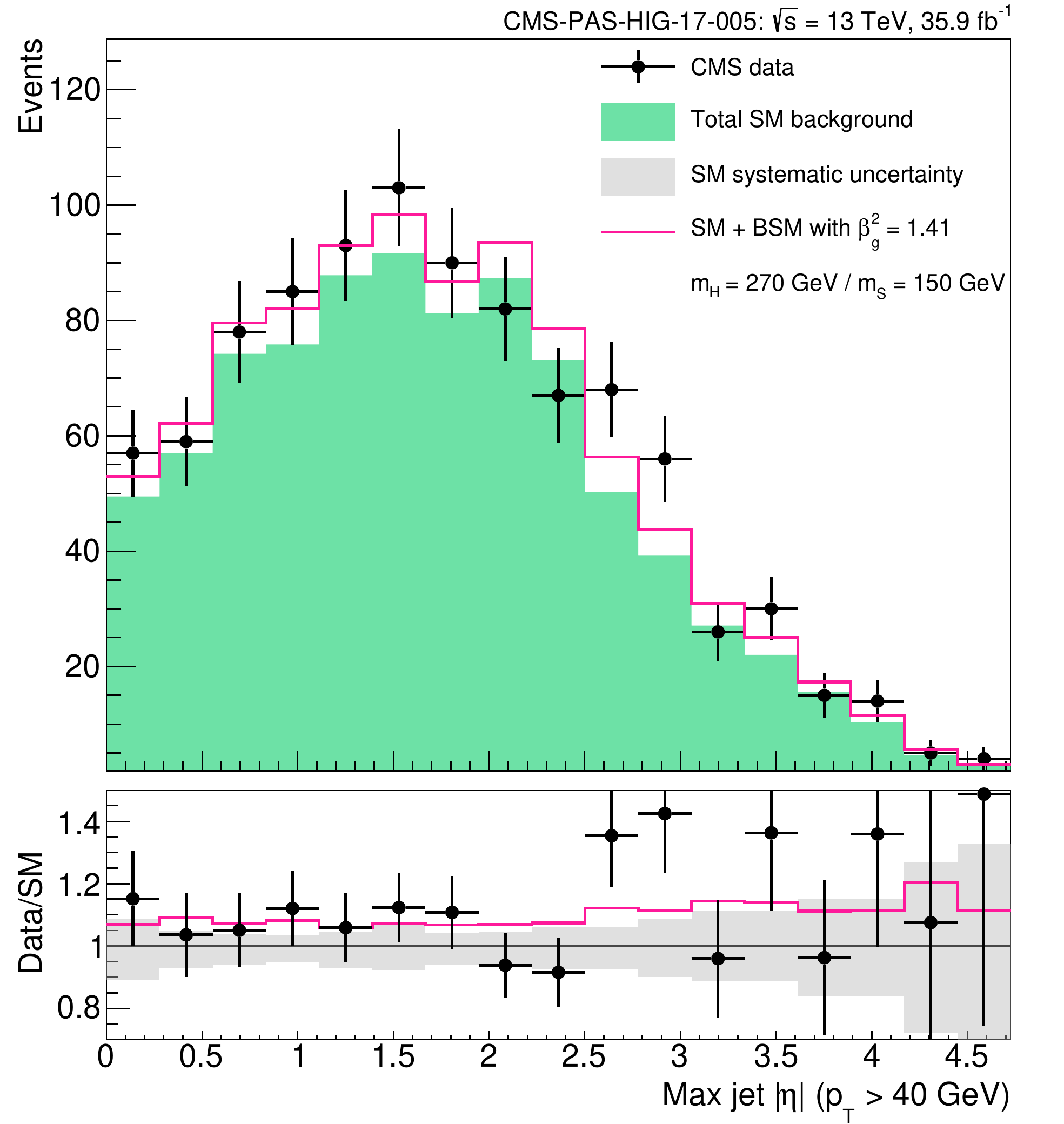}
    \includegraphics[width=0.49\textwidth]{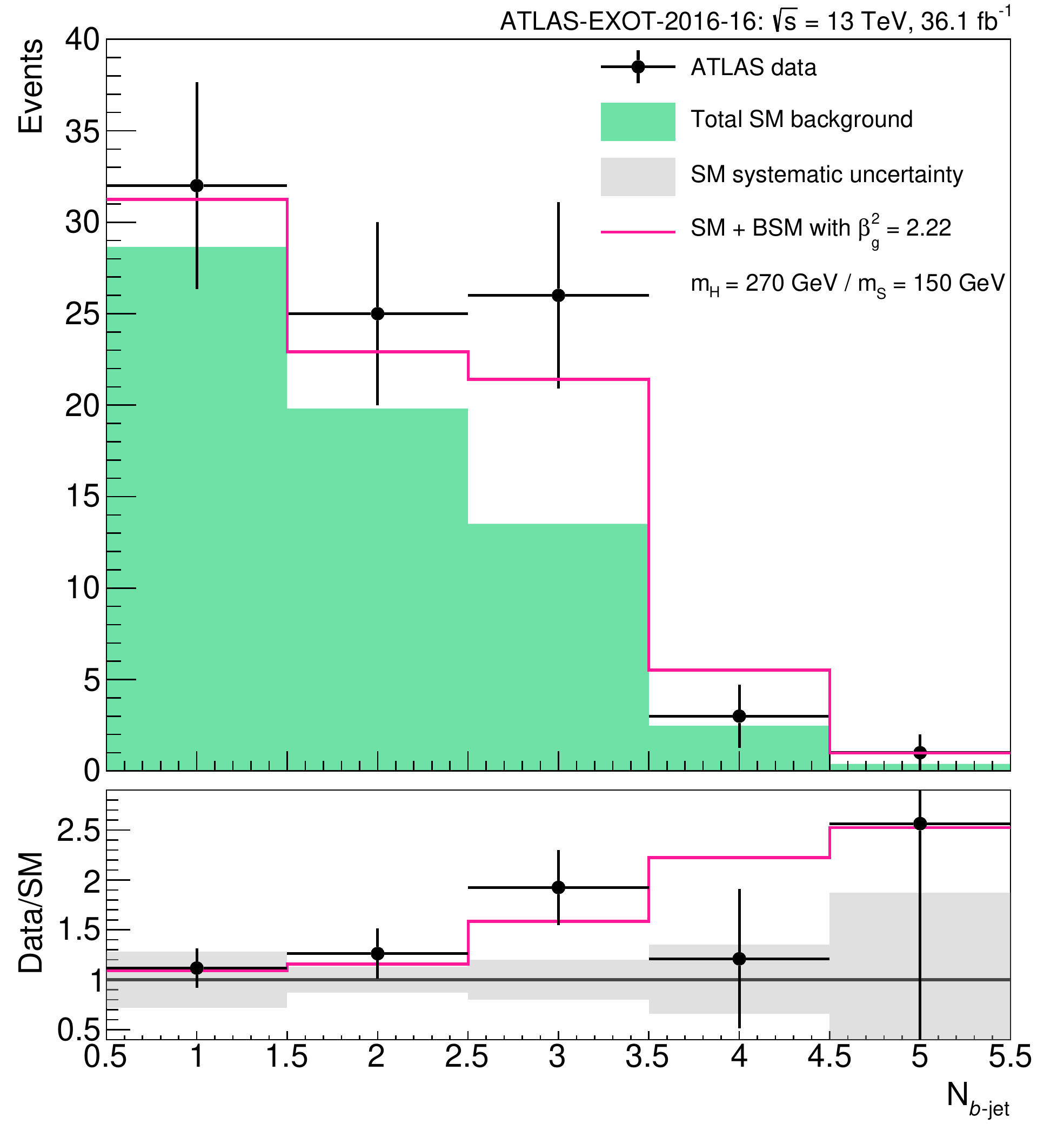}
    \caption{The SM+BSM fit results applied to searches for SS leptons in association with $b$-jets in the CMS (left) and ATLAS (right) Run 2 data sets.
    The discriminating variables in the fit are the highest value of pseudo-rapidity for high $p_\text{T}$ jets in the CMS search and the $b$-jet multiplicity in the ATLAS search.
    In each case, \bgs has been scaled to its best-fit value, while the SM predictions are shown at their nominal values without any fit constraints.
    The systematic uncertainties are shown as bin-by-bin effects, as reported by the experimental collaborations.
    However, in the fit an overall normalisation systematic uncertainty is considered for each measurement.
    }
    \label{fig:ssll_plots}
\end{figure}

The result of the fitting process is shown on the left in \Cref{fig:ssll_plots}.
Note that, although a bin-by-bin variation is shown as a systematic uncertainty in the plots in \Cref{fig:ssll_plots}, the fit makes use of an overall normalisation variation, including both the theoretical and experimental uncertainties calculated in the paper.
The resulting best-fit value of \bgs is $1.41\pm0.80$.
This corresponds to a significance of $Z=1.75\sigma$.
The overall agreement of the SM+BSM prediction compared with the data in the fit looks reasonable.
The addition of the BSM prediction helps to explain the overall elevation of the data compared with the SM, however this is not a significant effect due to the fact that this elevation can be covered by the systematic uncertainty.
A more interesting feature is the ability of the BSM prediction to partially explain the greater elevation for events with forward jets (that is, $|\eta|>2.5$).
The signal contribution in this region is dominated by the $tH$ and $ttH$ production mechanisms.
Due to the fact that the statistical uncertainty is still large for this measurement, it would be interesting to revisit this analysis with more data.

\subsection{ATLAS Run 2 search for same-sign leptons in association with \texorpdfstring{$b$}{b}-jets\label{sec:atlas_exot_2016_16}}

The Run 2 version of the ATLAS search for SS leptons in association with $b$-jets~\cite{Aaboud:2018xpj} provides a more statistically precise and systematically constrained measurement than that of Run 1 (the result discussed in \Cref{sec:atlas_exot_2013_16}).
The categorisation into SRs is slightly different, as in this case a selection of auxiliary plots show differential distributions combining all the SRs.
Therefore, the details of the SRs are not important in this study, but it should be noted that the combination of all SRs have minimal cuts of \MET~>~40~GeV and \HT~>~500~GeV.

Based on the signal to background ratio (using the MC predictions), it was clear that the most sensitive variable to the BSM prediction is that of the $b$-jet multiplicity.
In the ATLAS data, the distribution of $b$-jet multiplicity deviates from the SM in the bin with 3 $b$-jets.
The top associated production modes of the BSM scalar $H$ come with several $b$-jets in the final state, due to the decays of both the top quarks and the intermediate SM Higgs boson in the process.
The \ggf BSM production mode has a relatively small acceptance into the event selection of this measurement, and therefore the overall BSM prediction does a good job of explaining the excess in the distribution of $b$-jet multiplicity.
Similar to the CMS result in \Cref{sec:cms_pas_hig_17_005}, a single normalisation systematic uncertainty was considered for this measurement, corresponding to the overall systematic uncertainty on the SM background (including both theoretical and experimental effects).

The best-fit point for \bgs in the SM+BSM fit is found at $2.22\pm1.19$.
This corresponds to a significance of $Z=2.01\sigma$.
The result of this fit is shown in the distribution of $b$-jet multiplicity on the right of \Cref{fig:ssll_plots}.
While the significance of the fit is not particularly high, this is still an important result in this study.
This is because Ref.~\cite{Aaboud:2018xpj} provides the only distribution of $b$-jet multiplicity from ATLAS that can be related to the top associated production of the Higgs boson in multi-leptonic final states.
As demonstrated in Ref.~\cite{vonBuddenbrock:2017gvy}, the $ttH$ and $tH$ BSM production modes discriminate most strongly against the SM prediction of $b$-jet multiplicity, and this effect can be seen in the data here.
A similar (yet very strongly correlated) excess can be seen in the more recent ATLAS results in the search for four-top-quark production~\cite{ATLAS_4t}.

\subsection{ATLAS Run 2 ``top spin correlations'' \label{sec:atlas_conf_2018_027}}

Measurements of the azimuthal angle between OS leptons has historically been used to understand spin correlations in top quark decays.
The most recent study of the distribution, as presented by the ATLAS collaboration using a partial Run 2 data set~\cite{ATLAS-CONF-2018-027}, has shown that a significant deviation from the SM exists.
This result is interpreted by the experimental collaboration as being an indication of mis-modelled top quark spin correlations in the SM $t\bar{t}$ MC predictions.
In this article, we argue that the deviation is not necessarily due to the mis-modelling of the SM $t\bar{t}$ production process, but that a contamination of the proposed BSM signal studied here can alleviate the discrepancy between the data and the SM prediction.

In this article we consider the ``inclusive'' distribution of \dphill, the difference in azimuthal angle between different-flavour OS (DFOS) di-leptons. 
This selection requires that events have at least one $b$-jet.
The discrepancy can be seen in the detector level distribution, where in the data the lower values of \dphill are underestimated by the SM prediction and the higher values are overestimated.
However, it is also clear that there exists some uncertainty in the different SM MC predictions at high values, whereas the different MC predictions agree relatively well at low values.
Due to this discrepancy, we chose to use the \texttt{aMC@NLO} + \texttt{Pythia 8} prediction as a baseline, since it does the best job of describing the data in the region of high \dphill.
Thereafter, a conservative bin-by-bin systematic uncertainty was applied to cover the variation of the other SM $t\bar{t}$ predictions, constructed using the difference of the MC predictions tested in the ATLAS result.
This uncertainty varies from 0.3\% at $\dphill=0$ to around 2.5\% at $\dphill=\pi$.
In addition to this, all the appropriate experimental systematic uncertainties were applied to the SM predictions. 
The BSM contribution to the distribution is dominated by the \ggf production mode.
In general, the DFOS leptons come from the $S\to WW\to e\nu\mu\nu$ decays, whereas the extra $b$-jet(s) comes from the $h\to b\bar{b}$ decay mode.
Since the di-lepton pair comes from a cascaded decay via the heavy scalar $H$, the \dphill spectrum produced by the BSM \ggf production process peaks at low values, which is opposite to that of SM $t\bar{t}$ production.

\begin{figure}
    \centering
    \includegraphics[width=0.7\textwidth]{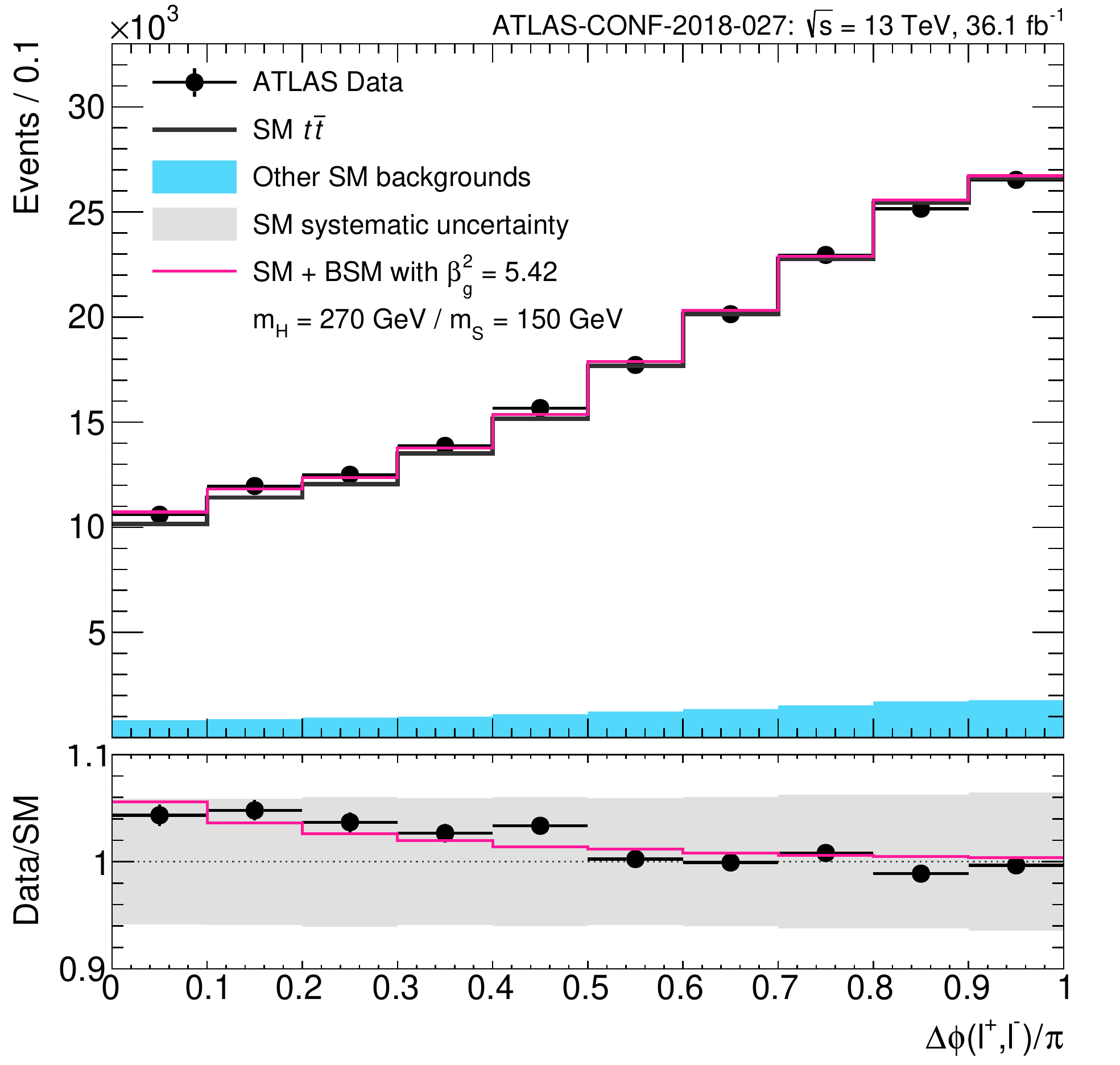}
    \caption{The SM+BSM fit result for the DFOS \dphill distribution reported by ATLAS using a partial Run 2 data set.
    Here, \bgs has been scaled to its best-fit value.
    The SM predictions and systematic uncertainties do not incorporate the constraints imposed by the fit.}
    \label{fig:atlas_conf_2018_027_plot}
\end{figure}

The SM+BSM fit does a remarkable job in describing the excess of data in the low end of the \dphill spectrum. The best-fit value of \bgs is slightly higher than most of the other fit results in this article, and is fit at $5.42\pm1.28$, corresponding to a significance of $4.06\sigma$.
The result of this fit can be seen in \Cref{fig:atlas_conf_2018_027_plot}. Note that, like all of the other data comparisons in this article, the systematic uncertainties and SM predictions are shown at their nominal and un-scaled values. It is only the BSM prediction that has been scaled to its best-fit normalisation. 
The inability of the SM to describe the data (even within systematic uncertainties) is due to the fact that the dominant systematic uncertainties only affect the overall normalisation of the SM, whereas the excess in the data clearly has a shape dependence. 
The BSM prediction matches this shape dependence very well, and hence the SM+BSM fit has a high significance.

We are impelled to comment on the top quark studies reported with a similar event selection and data set by the CMS collaboration~\cite{Sirunyan:2018ucr}. The CMS collaboration scrutinised a large number of hadronic and leptonic observables states and made comparisons to several MCs. The azimuthal angle between the two leptons was studied with \texttt{aMC@NLO} and \texttt{Pythia 8}, the MC that best describes it. However, while describing well \dphill, this MC is not able to adequately describe the transverse momentum of the $b$-jets. By contrast, the nominal MC used by ATLAS describes well the kinematics of the $b$-jets. Had the kinematics of the $b$-jets in the nominal MC used by CMS been re-weighted to the data one wonders if the description of \dphill would have deteriorated as observed by ATLAS. The conclusions of Ref.~\cite{Sirunyan:2018ucr} with regards to the study of \dphill differ from that drawn by ATLAS. That being said, CMS does acknowledge that a single MC is able to describe data in that significant disagreement is observed between the data and all predictions with regards to  several observables.

The impact of NNLO QCD corrections on the \dphill distribution has been reported in Ref.~\cite{Behring:2019iiv}. These results indicate that the discrepancy between the data and the NLO QCD-based MCs used in Ref.~\cite{ATLAS-CONF-2018-027} will be alleviated. This will also improve the ability of the simplified model considered here, as the signal normalisation obtained here with \dphill is elevated with respect to other results, such as those obtained in~\Cref{sec:cms_top_17_018} with a similar data set. 

\Cref{app:top} evaluates the impact of real gluon emissions on top of the QCD NLO matrix elements.  Additional gluon emission favours small \dphill while disfavouring small $m_{\ell\ell}$, where the bulk of the discrepancy is located. While additional gluon emissions improve the agreement in the \dphill distribution, it degrades the description of the di-lepton invariant mass. Whereas the significance observed on the basis of \dphill will decrease, the significance obtained with $m_{\ell\ell}$ distributions will be enhanced. A similar picture is appreciated when it comes to the production of non-resonant $W^+W^-$ production (see \Cref{sec:WW} and \Cref{app:top}). It is important to note that the ability of describing one observable undershoots the relevance of describing the kinematics of decay products in top processes simultaneously. One can never assume that by improving the description of one observable that others will also be described satisfactorily (see Ref.~\cite{Sirunyan:2018ucr}).

EW NLO corrections are available for $t\overline{t}$ production. These corrections are marginal in the corner of the phase-space where the discrepancies are observed~\cite{Denner:2016jyo}. EW effects become relevant for leptons with large \pT and large $m_{\ell\ell}$, regions of the phase-space that are distinct from the region of interest.

\subsection{ATLAS and CMS Run 2 measurements on \texorpdfstring{$WZ$}{WZ} production\label{sec:wz_results}}

Up until this point, this study has been concerned mostly with measurements that have dominant components relating to top quark production.
It is therefore fair to assume that the fit results up until this point might be biased towards potential mis-modelling of SM top production processes.
However, it can be shown that multi-lepton excesses exists also in measurements that are dominantly sensitive to electro-weak (EW) processes.

To demonstrate this, we have chosen to study the SM measurements of $WZ$ production as presented by the ATLAS~\cite{ATLAS-CONF-2018-034} and CMS~\cite{CMS-PAS-SMP-18-002} experiments in their Run 2 data sets.
These measurements both select events with exactly three leptons, two of which must be a same-flavour OS (SFOS) pair with a mass close to the $Z$ boson mass.
A cut on \MET is also made to select events containing a leptonically decaying $W$ boson.
The main difference between the ATLAS and CMS event selections is that CMS veto events containing $b$-jets, whereas ATLAS do not apply such a constraint.
The event selections applied to these searches are almost completely orthogonal to the other measurements considered in this article.

The only common distribution shown in the SR for both ATLAS and CMS is that of the $Z$ boson \pT (that is, the \pT of the SFOS di-lepton system with a mass closest to the $Z$ boson mass).
Since this variable relies only on the performance of reconstructing the momentum of light leptons, it is therefore relatively robust and not likely to suffer from theoretical mis-modelling.
For this reason, it was chosen to be the discriminating variable in the SM+BSM fit.
The SM prediction of the $Z$ boson \pT in the $WZ$ production process was calculated at next-to-leading order (NLO) in terms of quantum chromodynamics (QCD) corrections in both the ATLAS and CMS measurements, but not with the NLO EW corrections.
A study on the recent literature in SM $WZ$ production at the LHC has shown that the current predictions are relatively robust, with the overall NLO EW corrections having only a small effect on the $Z$ \pT spectrum~\cite{Biedermann:2017oae,Baglio:2018mhc}.

For these measurements, the \ggf BSM production mode again dominates over the top associated modes in terms of contamination into the SRs.
An MC study showed that the BSM prediction studied in this article seldom produces a SFOS lepton pair close to the $Z$ mass, and therefore the acceptance is still relatively low.
However, due to the fact that the $Z$ boson is most often produced through a cascaded off-shell decay (through the $h\to ZZ$ or $S\to ZZ$ decay mode), it has a very low \pT on average.
Therefore, it described the mild excesses seen in the ATLAS and CMS data at low $Z$ \pT relatively well.

\begin{figure}
    \centering
    \includegraphics[width=0.88\textwidth]{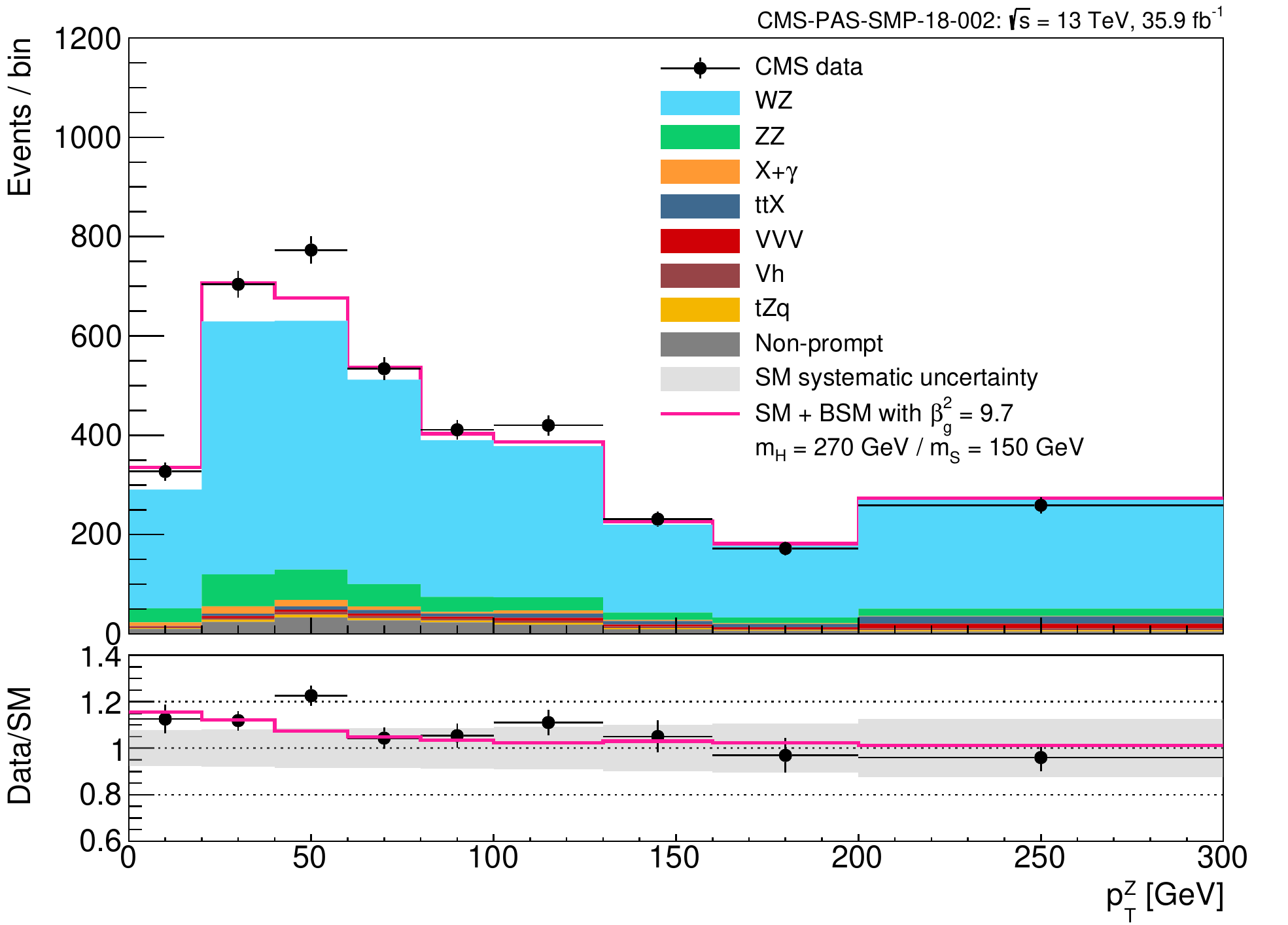}
    
    \includegraphics[width=0.88\textwidth]{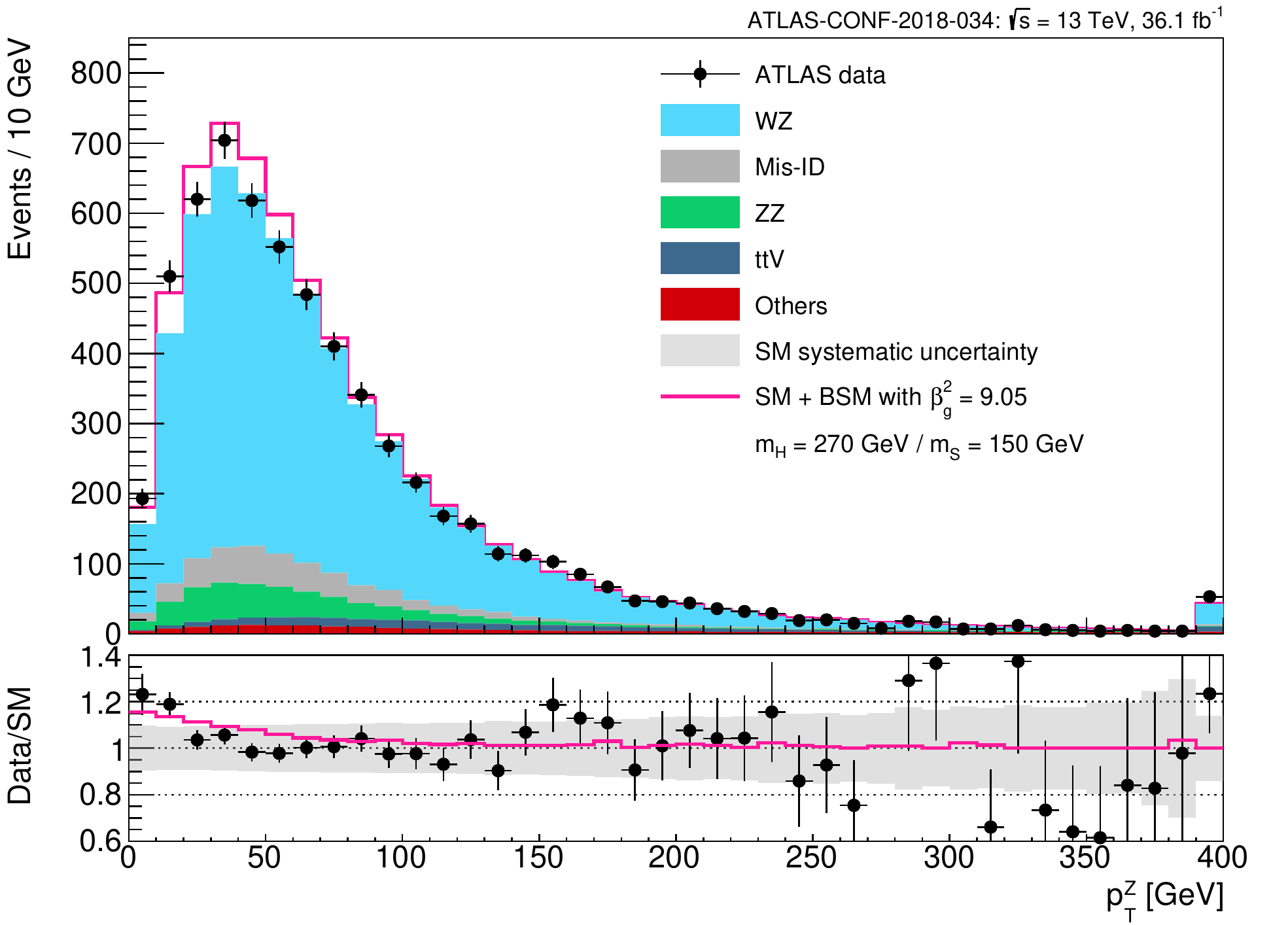}
    \caption{The SM+BSM fit result for the CMS (top) and ATLAS (bottom) measurements of SM $WZ$ production.
    Only the BSM prediction has been scaled to its best-fit value.
    The SM predictions and their associated uncertainties have been left at their nominal values.}
    \label{fig:wz_plots}
\end{figure}

It was decided that the $Z$ boson \pT would not be scaled to match the integral of the data in the tail of the distribution (which was done in the \mll distributions discussed in \Cref{sec:atlas_topq_2015_02} and \Cref{sec:cms_top_17_018}).
Therefore, the full set of experimental systematic uncertainties was applied to the SM distributions in the fitting procedure.
It was also evident that a comprehensive study on the theoretical uncertainties in the region of low $Z$ \pT for $WZ$ production should need to be understood so that the significance of the excess in the data could be characterised more accurately.
This was done in three different respects.
Firstly, a scale uncertainty as a function of the $Z$ \pT was derived by generating events in \texttt{POWHEG} at different combinations of 0.5, 1 and 2 times the renormalisation and factorisation scales, and thereafter identifying the largest deviation from the nominal scale.
This uncertainty was identified to be approximately a 5\% effect at \pT values lower than around 60 GeV, and up to a 10\% effect as the \pT becomes larger.
This effect was confirmed in events generated in \texttt{aMC@NLO}.
Secondly, the effect of changing the matrix element and parton shower of the event generation process was studied by comparing the $Z$ \pT for all the combinations of events generated with \texttt{POWHEG} and \texttt{aMC@NLO} and showered with \texttt{Pythia 8} and \texttt{Herwig 7}~\cite{Bellm:2017bvx}.
From this study it was determined that \texttt{POWHEG} events showered with \texttt{Pythia 8} does the most reasonable job in explaining the entire spectrum. 
The \pT spectrum produced by \texttt{aMC@NLO} was relatively hard, and would therefore enhance the excess seen in the data.
The most conservative choice was to use \texttt{POWHEG}+\texttt{Pythia 8} as the nominal prediction.
Finally, the effects of changing parton density function (PDF) sets was studied by producing the $Z$ \pT spectrum with the \texttt{CT14}~\cite{Dulat:2015mca}, \texttt{MMHTnlo}~\cite{Harland-Lang:2014zoa} and \texttt{PDF4LHCnlo}~\cite{Butterworth:2015oua} PDF sets.
This effect was seen to only alter the normalisation of the prediction slightly, and therefore was not considered a significant systematic uncertainty.
Several theoretical studies on the $Z$ \pT spectrum can be found in \Cref{app:wz_theory}.

In both the ATLAS and CMS measurements the SM+BSM fit favoured the best-fit value of \bgs to be rather high, at $10.65\pm3.24$ for the ATLAS measurement and $10.28\pm3.76$ for the CMS measurement.
This corresponds to significance values of $3.00\sigma$ and $2.51\sigma$, respectively.
The distributions with the BSM prediction scaled to the best-fit value can be seen in \Cref{fig:wz_plots}.
Note that the ATLAS result may appear to have been incorrectly fit, since the BSM is scaled above most of the data points.
However, it appears this way because the SM predictions and their associated systematic uncertainties have been left at their nominal values.
Within the systematic uncertainty, the SM prediction is pulled to a lower normalisation in order to explain the depletion of events in the tail of the distribution.

\subsection{Combination\label{sec:combination}}

\begin{table}
    \centering
    \begin{tabular}{l|c|c}
    \hline
        \multicolumn{1}{c|}{\textbf{Selection}} & \multicolumn{1}{|c}{\textbf{Best-fit} \bgs} & \multicolumn{1}{|c}{\textbf{Significance}}  \\
        \hline
        ATLAS Run 1 SS leptons + $b$-jets & $6.51\pm2.99$ & 2.37$\sigma$  \\
        ATLAS Run 1 DFOS di-lepton + $b$-jets & $4.09\pm1.37$ & 2.99$\sigma$ \\
        ATLAS Run 2 SS leptons + $b$-jets & $2.22\pm1.19$ & 2.01$\sigma$ \\
        CMS Run 2 SS leptons + $b$-jets & $1.41\pm0.80$ & 1.75$\sigma$ \\
        CMS Run 2 DFOS di-lepton & $2.79\pm0.52$ & 5.45$\sigma$ \\
        ATLAS Run 2 DFOS di-lepton + $b$-jets & $5.42\pm1.28$ & 4.06$\sigma$ \\
        CMS Run 2 tri-lepton + $\MET$ & $9.70\pm3.88$ & 2.36$\sigma$ \\
        ATLAS Run 2 tri-lepton + $\MET$ & $9.05\pm3.35$ & 2.52$\sigma$ \\
        \hline
        Combination & $2.92\pm0.35$ & 8.04$\sigma$ \\
        \hline
    \end{tabular}
    \caption{A summary of the SM+BSM fit results for each measurement considered in this article, along with the result of their combination.}
    \label{tab:combination_list}
\end{table}

\begin{figure}
    \centering
    \includegraphics[angle=90,width=0.8\textwidth]{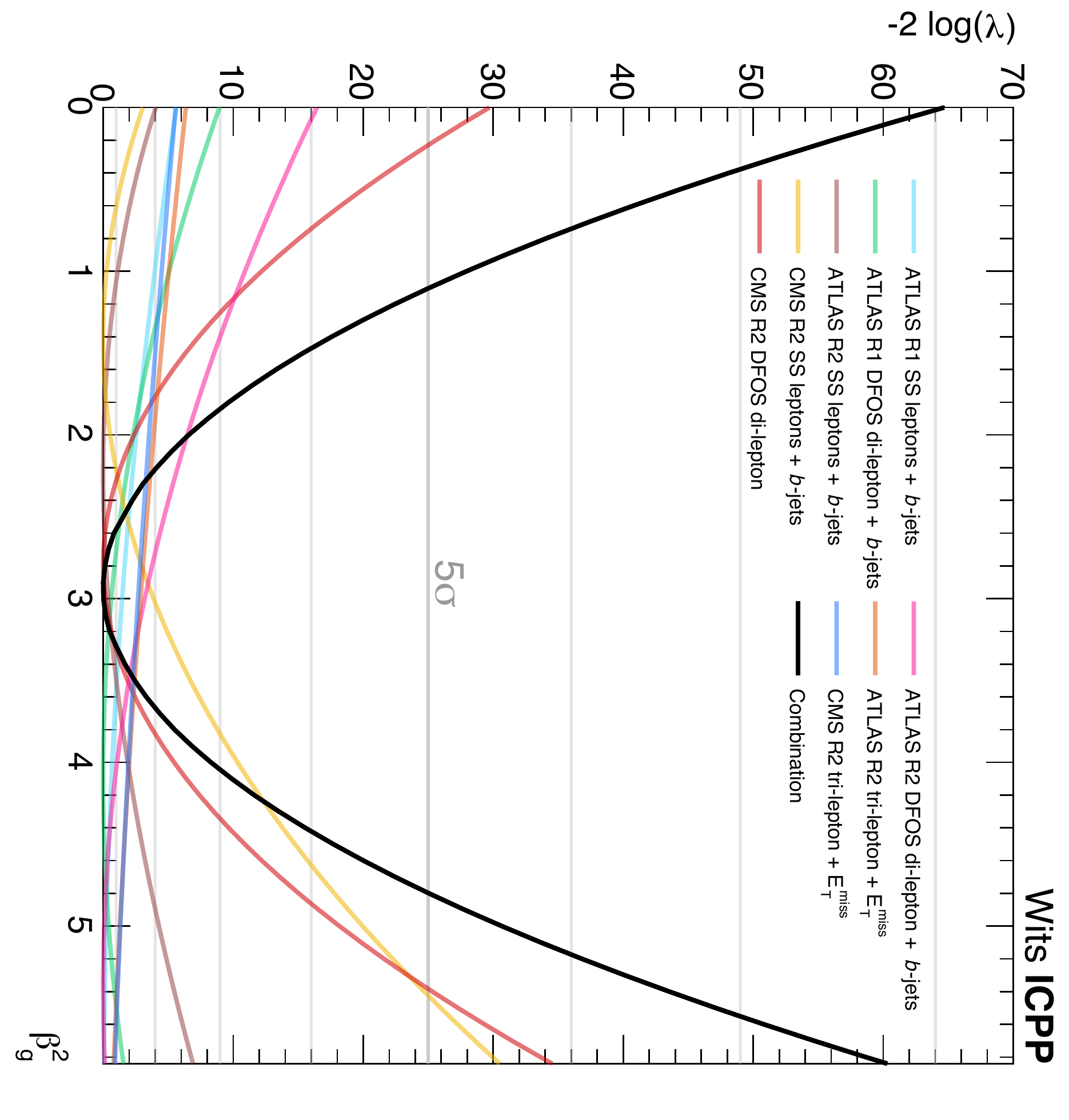}
    \caption{The (negative 2 times the logarithm of the) profile likelihood ratios for each of the individual fit results considered in this article, overlaid with that of their combination.
    The significance of a result is calculated as the square root of the point which intersects the $y$-axis (that is, the SM-only hypothesis where $\bgs=0$).}
    \label{fig:plr}
\end{figure}

Each of the results studied in this article make use of a profile likelihood ratio to constrain the single fit parameter \bgs under an SM+BSM hypothesis.
With these profile likelihood ratios constructed as a function of \bgs, it is relatively straightforward to perform a simultaneous fit on all of the results considered and therefore make a combination of the independent data sets under the SM+BSM hypothesis.
The combined profile likelihood is constructed by multiplying the profile likelihood ratios for each individual measurement.
Then, the best-fit value of \bgs and significance can be calculated similarly to the individual results (i.e. by minimising \Cref{eqn:plr} and using \Cref{eqn:significance}).
Doing so constrains the parameter \bgs to the value $2.92\pm0.35$, which corresponds to a significance of $Z=8.04\sigma$ in favour of the SM+BSM hypothesis over the SM-only hypothesis.
A summary of all the individual fit results, as well as the combination, can be seen in \Cref{tab:combination_list}.
In addition to this, each of the individual profile likelihood ratios are shown in \Cref{fig:plr}, with the combined case shown in black.

The significance for each individual result and the combination is calculated assuming one degree of freedom.
This is because, as it was stated in \Cref{sec:intro,sec:model}, the masses of the model were fixed from previous un-correlated studies, and the couplings of $S$ and $H$ were fixed so as to benchmark the model in a transparent way.
It is not clear from this study if allowing extra degrees of freedom in the fit would increase or decrease the significance, since another degree of freedom may have found a better best-fit point, but the significance would have to be calculated in two dimensions (thus suppressing the result).
However, the important point to note is that the apparent discrepancy with respect to the SM is significantly explainable with only one degree of freedom and over an ensemble of results that involve processes with vastly different production rates.
This fact must be highlighted as BSM models become more complicated in the search for new physics.

In terms of the combination that we have performed, the calculated significance can only be treated as an estimate of the ``true'' value, due to the fact that we lack much of the necessary information to formally combine the results.
For one, statistical correlations for the effects of systematic uncertainties have not been accounted for.
Having said this, since there exists a diverse set of measurements in the combination, it is not obvious that incorporating such correlations would have a big effect on the final fit result.
Another necessary bit of information that we lack is the exact way in which the experimental systematic uncertainties affect the shapes of the SM distributions that have been fit.
The best possible approximation that could have been made was to incorporate bin-by-bin variations for those that we do have an understanding of. However it stands to reason that a more rigorous and insightful application of systematic uncertainties to the distributions could change the final fit results (although not significantly).

Looking at \Cref{tab:combination_list}, it is apparent that the individual best-fit values seem to have a large spread.
One can quantify this spread by looking at the compatibility of each pair of results using the statistical significance of their difference.
For any two results $i$ and $j$, the compatibility can be quantified as: 
\begin{equation}
    C=\left|\mu_i-\mu_j\right|/\sqrt{\sigma_i^2 + \sigma_j^2},
    \label{eqn:compatibility}
\end{equation}
where the denominator is the difference of the best-fit values of \bgs and the denominator is their combined standard deviation.
The largest incompatibility is between the CMS Run 2 SS leptons + $b$-jets and the ATLAS Run 2 DFOS di-lepton + $b$-jets, with $C=2.66\sigma$.
The two results that drive the discrepancy (i.e.~the CMS Run 2 DFOS di-lepton and the ATLAS Run 2 DFOS di-lepton + $b$-jets) are indeed relatively compatible, with $C=1.90\sigma$.
All physics discussion relating to these results is presented in \Cref{sec:discussion}.

Since many of the results that were studied contain event yields that are extremely statistically precise, it is obvious that systematic uncertainties play a dominant role in the determination of the significance of the SM+BSM fit. 
For each result the profile likelihood ratio compared with the negative log likelihood (NLL) scaled to the same minimum can be seen in \Cref{app:syst}. 
Comparing the two functions is an appropriate way to gauge how the systematic uncertainties imposed in the fits affects the overall significance of the fit.

\subsection{ATLAS and CMS non-resonant \texorpdfstring{$W^+W^-$}{W+W-} differential distributions}
\label{sec:WW}

Discrepancies between data and MC in the measurement of the differential distributions were reported by ATLAS and CMS with Run 1 data and pointed out in Ref.~\cite{vonBuddenbrock:2017gvy}. Final states considered here comprise two high \pT electrons or muons with a full hadronic jet veto. The latter is intended to suppress the contribution from top backgrounds. The bulk of the discrepancy occurs here for $m_{\ell\ell}<100$~GeV, as seen elsewhere.  

Recent results reported by the ATLAS collaboration confirm with more statistics the anomalies described above in di-lepton final states~\cite{Aaboud:2019nkz}, where QCD NNLO corrections have been applied to $q\overline{q}\rightarrow W^+W^-$ production ~\cite{Gehrmann:2014fva,Grazzini:2016ctr,Hamilton:2016bfu,Re:2018vac}, QCD NLO corrections to non-resonant $gg\rightarrow W^+W^-$~\cite{Caola:2015rqy} and EW NLO corrections~\cite{Biedermann:2016guo}. The event selection used in Ref.~\cite{Aaboud:2019nkz} uses thresholds on leptons and jets that are somewhat more elevated compared to those used in Run 1. The impact of these on the lepton kinematics was studied here, where compatibility between Run 2 and Run 1 results was verified.

All in all recent results regarding the non-resonant $W^+W^-$ production indicate that the discrepancies between the data and the SM prediction for $m_{\ell\ell}<100$~GeV seen since Run 1 are unlikely to be due to statistical fluctuations. This final state is not added to the combination performed here for technical reasons. That said, these discrepancies with Run 1 data were also interpreted with the simplified model used here in Ref~\cite{vonBuddenbrock:2017gvy} with results that are similar to those obtained here. 

\section{Discussion\label{sec:discussion}}

It goes without saying that the large combined significance discussed above should be subject to some criticism.
From a statistics standpoint, a common issue raised in such circumstances is that of a look elsewhere effect.
Traditionally, a look elsewhere effect will suppress the significance of a fit result that was performed on an unexpected (or un-predicted) deviation from the null hypothesis.
This has the advantage of reducing cognitive bias in terms of model building, such that it is a safeguard against tuning a model's parameters in order to describe a fluctuation.
In terms of the fits performed in this article, we strongly believe that a look elsewhere effect is not appropriate.
As discussed in \Cref{sec:intro,sec:model}, the mass points in this article were decided to be fixed \textit{a priori}, and therefore the model was not tuned to explain the data.
In truth, a mass scan would be interesting to study, but this is left for a future work.

The measurements against which the model was fit were decided based on the predicted signatures of the model as described in previous studies~\cite{vonBuddenbrock:2016rmr,vonBuddenbrock:2017gvy}.
The ensemble of measurements we constructed is an exhaustive set of such measurements currently available, taking into account that adding additional measurements could account for the double counting of data sets.
One might ask why such a prominent excess should show up in SM measurements instead of those that aim to discover new physics, particularly in multi-lepton final states (with or without $b$-jets).
BSM searches for such final states are mostly conducted in a SUSY framework, where multiple leptons are often produced in cascade decays of charginos and neutralinos.
The reason that the model discussed in this article is not sensitive to such searches is that a key signature in SUSY searches is the requirement of large values of \MET, \HT, or the invariant mass of various multi-lepton systems.
These cuts are designed for searches in a phase space well above the EW scale, whereas our model deals with particles in the neighbourhood of the EW scale.
Due to the fact that the model discussed in this article produces its final state via off-shell cascaded decays, the produced leptons and $b$-jets tend to be far softer than those produced in SUSY processes, and therefore the predicted phase space tends to be quite different than what the typical SUSY searches require.
This is not to say that there is no overlap at all.
A recent detailed study by the GAMBIT collaboration has shown that, in a relatively generic and model independent way, an excess with a local significance of $3.5\sigma$ can be deduced in a statistical combination of various SUSY results from the LHC and the Large Electron-Positron Collider~\cite{Athron:2018vxy}.
By and large, the results they have considered relate to the production of multiple leptons.
It stands to reason that the excess they have calculated is in some way correlated with the excess we have presented here.
However, determining the extent of this is beyond the scope of this article.

There are several issues that deserve some attention with regards to the BSM model considered in this article.
With the single degree of freedom \bgs, the BSM model is able to shed light on the magnitude of the excesses in the data.
However, it is clear from \Cref{tab:combination_list} that in the fits there exists a tension between different final states.
Their statistical compatibility has already been discussed in \Cref{sec:combination}, however some further considerations should be noted.
In particular, the fit results for the $WZ$ measurements seem to exclusively require more BSM signal events than any other measurement.
Interestingly enough, the ATLAS and CMS results are remarkably consistent with one another, having $C=0.13\sigma$ (see \Cref{eqn:compatibility}).
However, this strong compatibility is not shared with the rest of the ensemble, where for  the $WZ$ results \bgs is over a factor of 3 larger than for most of the other results.
Apart from the $WZ$ results, the rest of the ensemble exhibits a noticeable spread around the combined best-fit mean value.
While we do argue that the results are statistically compatible, the spread could be an indication of underlying effects that are not considered in this article.
Most notably, \bgs for the ATLAS Run 2 OS (spin correlation) measurement is almost a factor of 2 larger than the best-fit value.
If one is to believe that the excesses in data truly are the result of new physics processes at the LHC, then what can be said is that the BSM model used in this article does not predict the correct relative mixture of events in terms of lepton and $b$-jet multiplicity.
The simplified assumption of only one degree of freedom appears to be incapable of constructing a coherent prediction that is able to concurrently explain all of the excesses discussed in this article.
As mentioned above, one interesting avenue to explore is multiple mass points of the BSM mode.
Due to the sensitivity of the mass dependent BRs for the Higgs-like scalar $S$, a wide range of $b$-jet and lepton multiplicities could be explored. 
Studies of top processes perfomed here indicate that leptonic observables, such as the di-lepton invariant mass and the transverse mass of the  system made by the di-lepton and missing transverse energy, depend weakly on the $b$-tagged and light jet multiplicities. 
In this light, the need to present di-lepton results as a function of the jet multiplicity is emphasised. 

It should be re-stated that if new physics is indeed responsible for the excesses discussed in this article, the simplified BSM model we have presented is not an ideal candidate to explain them consistently. 
This statement is made stronger by the inability for the simplified model to describe the localised excess seen in the ATLAS Run 2 $m_\text{T}$ distribution (see \Cref{fig:atlas_mT}). 
In addition to a mass scan for $S$ and $H$, it would be worthwhile looking into different assumptions for the decays of $S$. 
After all, allowing the $S$ to have Higgs-like BRs was a convenient assumption to begin with, only because it assists in reducing the number of degrees of freedom in the model. 
An interesting alternative could be to study the heavy neutrino model introduced in Ref.~\cite{vonBuddenbrock:2017gvy}, since many of the kinematic distributions are similar to the model we have used here, and the extra degrees of freedom may make for a better fit (albeit possibly not as significant).
Having said this, the BSM model used in this article does a remarkable job given that it requires only one degree of freedom. 
In addition, the possible shortcomings of the simplified model discussed here tend to decrease the significance reported here. 
In other words, the significance reported here is reflective of the degree of discrepancy between the data and the SM MC to the extent it is described by the model. 
Therefore, the magnitude of the discrepancy between the data and the MC may well be greater than the significance reported here. 

An important input to the possible BSM interpretation considered here is the search for $H\rightarrow 4W$ performed by the ATLAS collaboration~\cite{Aaboud:2018ksn} with data taken in 2015 and 2016. 
Whereas the anomalous production of OS, SS and three leptons in particular corners of the phase-space seems to be consolidating, Ref.~\cite{Aaboud:2018ksn} has not reported an excess. 
Within the simplified model considered here, this result could suggest that $m_H<m_S+m_h$, where the expected yield for four high transverse momentum leptons decrease rapidly with decreasing $m_H$. 
Another argument in favour of $m_H<m_S+m_h$ lies in the observed rate of the SM Higgs boson. 
The measured rate of $h$ is about 10-15\% greater than that predicted in the SM. 
The value of \bgs determined here would lead to an excess in the rate of $h$ of about twice the size. 
While the tension is not yet strong, it can be alleviated by $m_H<m_S+m_h$. 
Here $h$ would decay off-shell half the time, thus reducing the yield of on-shell $h$ production in critical decays, such as $h\rightarrow\gamma\gamma,ZZ\rightarrow 4\ell$.
The absence in the data of a signal $S\rightarrow ZZ\rightarrow 4\ell$ also seems to indicate that the simplified model used here displays too a naive implementation of the couplings to weak bosons or that the lepton anomalies are not mediated by weak bosons in the SM. 

Another interesting prospect would be to consider a VBF production mode for $H$. It was stated in \Cref{sec:model} that a small effective $H$-$W$-$W$ coupling, which has the advantage of enhancing the single top associated production mode, would suppress VBF.
Throughout the study, however, it was determined that the single top associated production mode of $H$ does not have a significant impact on the fit results for all of the considered measurements.The only non-negligible contributions of the $tH$ production mechanism for the measurements considered in this article are those that search for SS leptons in association with $b$-jets. Removing the $tH$ contribution to the signal could have as much as a 20\% effect on the best-fit values for \bgs in these measurements. However, they are not very sensitive measurements, and $tH$ has a negligible impact on the most sensitive measurements studied in this article. Therefore, the assumption on the $H$-$W$-$W$ coupling could indeed be relaxed, which would open up new possibilities of measurements that could be probed. Still more interesting would be to explore different models of the $H$ and $S$ bosons themselves, in terms of different possible spins and decay modes. This is left for future studies.

QCD NNLO corrections in di-lepton final states emerging from top backgrounds 
are not expected to change the conclusions of this paper. EW NLO corrections are marginal in the region of the phase-space of interest. Recent results regarding the non-resonant $W^+W^-$ differential cross-sections confirm the discrepancies seen between the Run 1 data and the SM MC at low values of $m_{\ell\ell}$. These discrepancies can also be described by the simplified model used here. The leptonic kinematics of the region of the phase-space where the data deviates from SM MCs in non-resonant $W^+W^-$ and top processes are similar. These discrepancies can also be described by the simplified model used here.

Should the $8.04\sigma$ significance of the combined fit stand the test of time and scientific criticism, it will present a challenge to our current understanding of physics at the LHC. Theoretically, it appears that the current set of tools used to describe SM processes is failing to do so, even in measurements of quantities as simple as the momenta of leptons. Whether or not this failure is due to BSM physics at the LHC remains to be seen. Any contamination of BSM physics relating to the Higgs sector would have profound impacts on the measurements of its mass and couplings. In any event, it is necessary to try and understand the data with as little bias as possible as we strive to solidify our understanding of the SM and beyond.

\acknowledgments

The authors would like to thank Ansgar Denner and Alexander Savin for useful discussions. 
The authors are grateful for the support from the DST and NRF through the SA-CERN consortium. 
SvB acknowledges the National Institute for Theoretical Physics (NITheP) for funding and support. 
AFM is grateful for funding via the CAS-TWAS President's Fellowship. 
YF and AFM are partially supported by the Beijing Municipal Science and Technology Commission No.Z181100004218003.

\appendix

\section{Studies of \texorpdfstring{$t\overline{t}$}{tt bar} and \texorpdfstring{$tW$}{tW} processes\label{app:top}}

In this section we succinctly describe the checks performed to estimate variations in di-lepton distributions emerging from $t\overline{t}$ and $tW$ processes. These variations are evaluated with QCD NLO calculations embedded in MCs with the \texttt{POWHEG}  methodology~\cite{Frixione:2007nw,Re:2010bp,Jezo:2016ujg}.
Studies performed here are focused on two critical observables: the di-lepton invariant mass and the transverse mass of the di-lepton system and the missing transverse energy (see~\Cref{sec:atlas_topq_2015_02} and~\Cref{sec:cms_top_17_018}). 

Studies are performed in a fiducial region defined below that includes the presence of one electron and one muon of opposite charge that pass the following requirements:

\begin{itemize}
	\item Electrons need to have $|\eta|<2.47$, excluding $1.37 < |\eta| < 1.52$.
	\item Muons need to have $|\eta|<2.5$.
	\item The leading and sub-leading lepton need to have $p_\text{T}>22$~GeV and $p_\text{T}>15$~GeV.
	\item The di-lepton invariant mass needs to be $\mll > 10$~GeV and $\MET>20$~GeV
	\item The largest of the transverse masses of a lepton and the $\MET$ has to be greater than 50~GeV.
\end{itemize}

\begin{figure}
    \centering
    \includegraphics[width=0.48\textwidth]{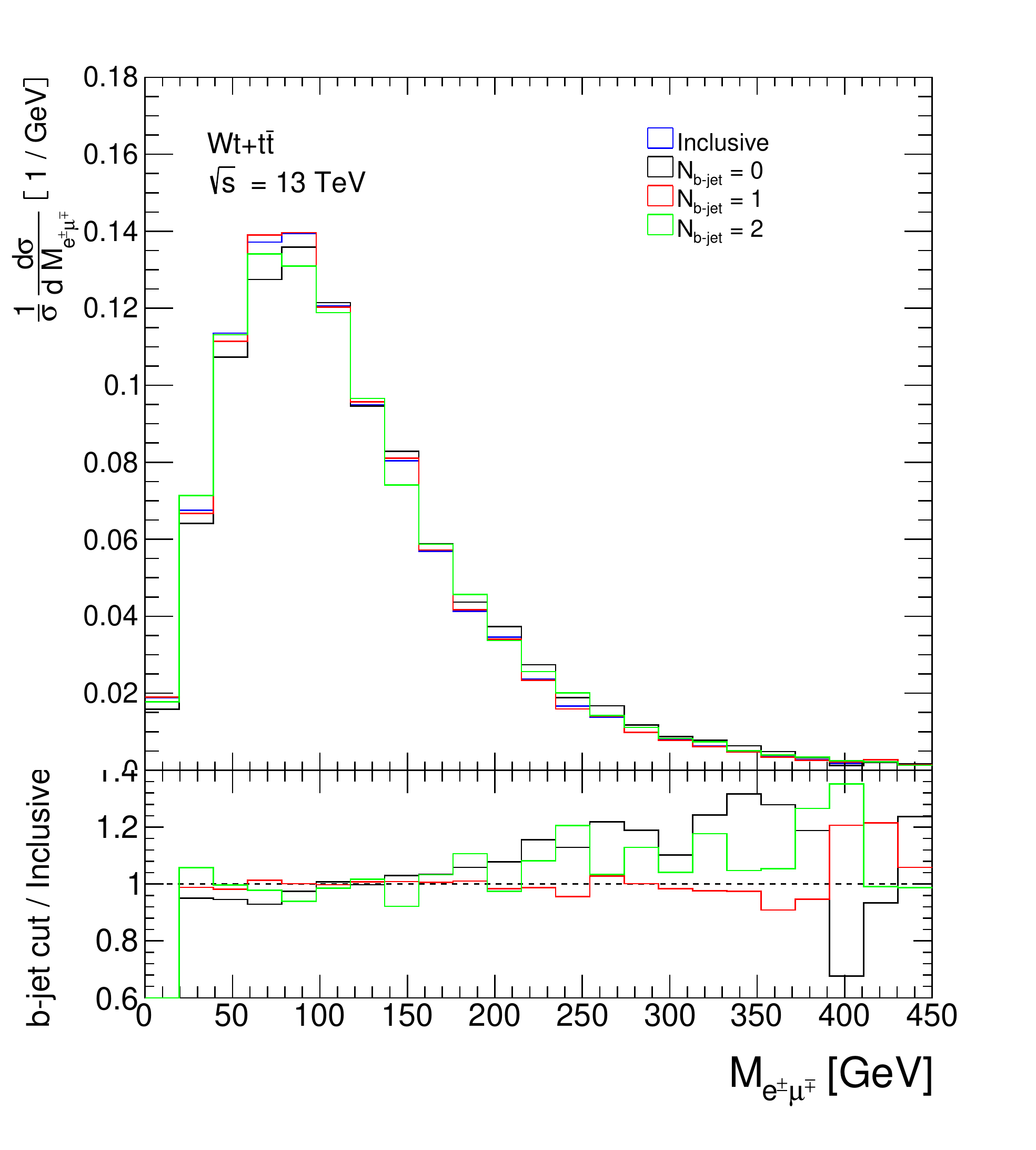}
    \includegraphics[width=0.48\textwidth]{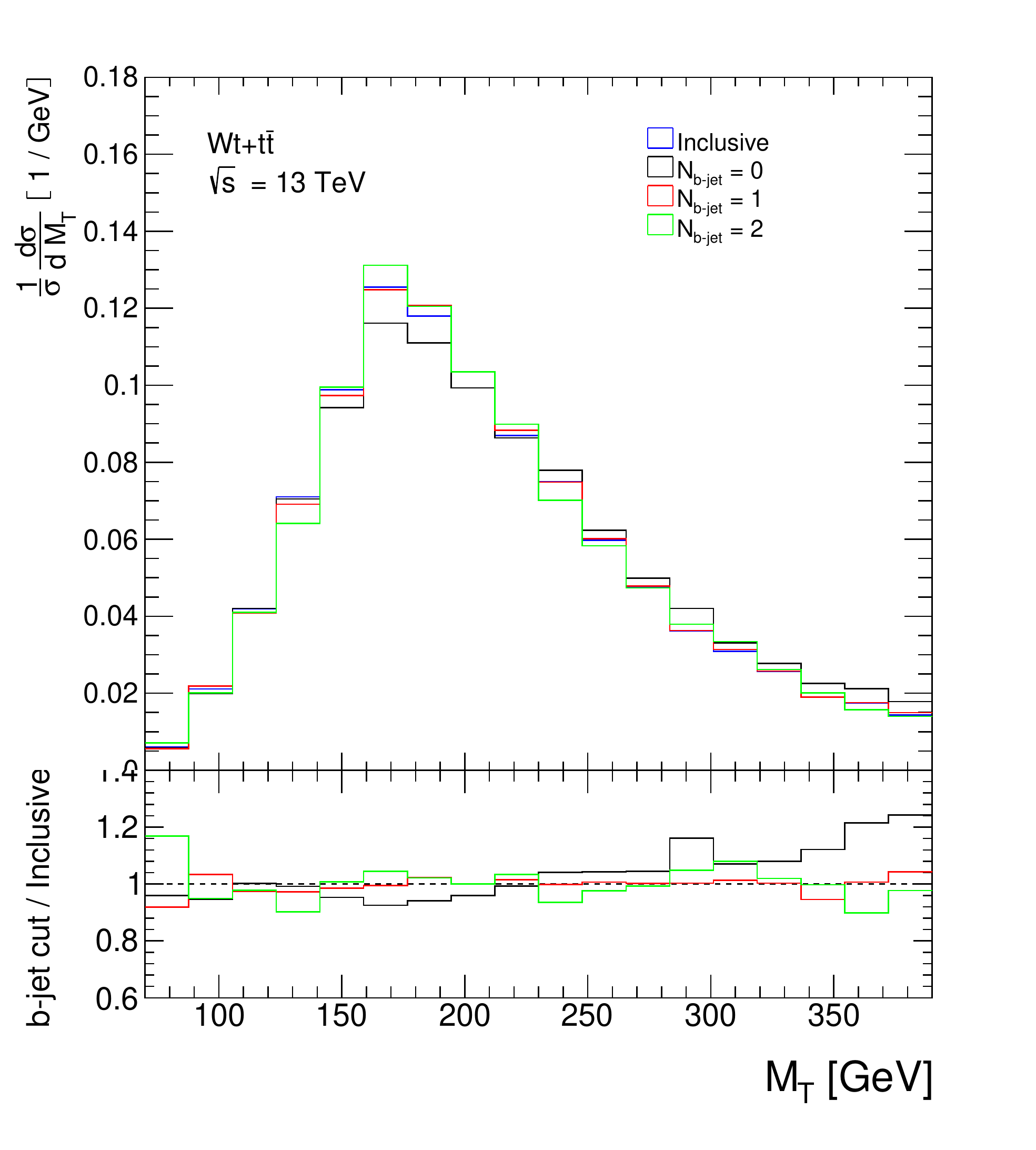}
    \caption{Leptonic distributions produced by $t\overline{t}$ and $tW$ processes (see text) as a function of the $b$-tagged jet multiplicity. The di-lepton invariant mass (left) and the transverse mass of the di-lepton and missing transverse energy system are displayed. Distributions are normalised to unity. The insert shows the ratio of the distributions with exclusive $b$-tagged jet bins relative to that obtained inclusively.}
    \label{fig:mllmt_bjet}
\end{figure}

Hadronic jets are defined as being within $|\eta|<4.5$ and with $p_\text{T}>20$~GeV, where $b$-tagged jets are defined as in~\Cref{sec:results}. The relative contribution of $tW$ with respect to $t\overline{t}$ of the final state considered here depends on the $b$-tagged and jet multiplicity in general.\footnote{The relative contribution of $tW$ and other processes in final states with one electron and one muon is nicely illustrated in Figure 3 of Ref.~\cite{Sirunyan:2018lcp}.} It is therefore relevant to scrutinise the dependence of di-lepton quantities as a function of the $b$-tagged jet multiplicity. \Cref{fig:mllmt_bjet}  displays the di-lepton invariant mass (left) and the $m_\text{T}$ (right) for different exclusive $b$-tagged jet multiplicities. The distributions for exclusive $b$-tagges jet bins are compared to that obtains inclusively. One appreciates that the leptonic variables shown in~\Cref{fig:mllmt_bjet} very weakly with the $b$-tagged jet multiplicity, where the transverse mass does not change appreciably. The di-lepton invariant mass changes less than 5\% for $\mll<200$~GeV whereas for greater values differences remain within 10\%.

In order to draw a quantitative statement with regards to the robustness of the leptonic quantities considered here, two ratios are constructed that are geared towards quantifying uncertainties associated with \Cref{fig:mll_plots} and \Cref{fig:atlas_mT}: 
\begin{equation}
    R(\mll) = \frac{\int_0^{100}{\frac{d\sigma}{d\mll}}d\mll}{\int_{110}^{\infty}{\frac{d\sigma}{d\mll}}d\mll},
\end{equation}

\begin{equation}
    R(m_\text{T}) = \frac{\int_0^{200}{\frac{d\sigma}{dm_\text{T}}}dm_\text{T}}{\int_{210}^{\infty}{\frac{d\sigma}{dm_\text{T}}}dm_\text{T}}
\end{equation}

A first check is performed by varying the renormalisation and factorisation scales by multiplying and dividing by a factor of two. 
The variation of $R(\mll)$ and $R(m_\text{T})$ remain within 2\% and 1\%, respectively. 
If the factor is increased to 3, the variation is less than 5\% and 2\%, respectively. 
The ratios are evaluated with different parton density function parameterisations where a variation of 2.5\% is estimated. 
The tighter event selection used to obtain \Cref{fig:atlas_mT}, where exactly one $b$-tagged jet without additional hadronic jets is required, is also applied. 
The value of $R(\mll)$ and $ R(m_\text{T})$ change by less than 2\% and 1\%, respectively. 
This speaks to the robustness of the leptonic quantities with respect to the $b$-tagged jet and light jet multiplicities. 
Scale variations are also performed after the requirement of the tighter event selection on the hadronic final state. 
The results are similar to those obtained inclusively. 

Checks are made with the two different schemes used to merge $t\overline{t}$ and $tW$ processes: Diagram removal and Diagram subtraction~\cite{Re:2010bp,Frixione:2008yi}. The differences observed in $R(\mll)$ and $R(m_\text{T})$ due to the change of scheme is well below 1\%.

The uncertainties obtained here with regards to the di-lepton invariant mass are consistent with studies reported by the collaborations (see \Cref{fig:mll_plots} and corresponding text). These uncertainties are not used for the statistical study performed here, where those reported by the experiments are used instead.

These studies with regards to the transverse mass seem to suggest that with current tools it is improbable to cover the differences between data and MC observed in \Cref{fig:atlas_mT} with existing tools and standard procedures. We have not performed a systematic study of differences between various generators for this particular corner of the phase-space. As a matter of fact, no statistical statement is drawn from \Cref{fig:atlas_mT} and this discrepancy is not included in the combination reported in~\Cref{sec:combination}. However, given that significant discrepancies are emerging in opposite sign charged leptons in association with $b$-tagged jets, as detailed in  \Cref{sec:atlas_topq_2015_02},   \Cref{sec:cms_top_17_018} and \Cref{sec:atlas_conf_2018_027}, it seems advisable that the experiments scrutinise the differences between data and MC as a function of $b$-tagged jet multiplicity. This is in contrast to measurements of total and differential cross-sections pertaining top and not resonant $WW$ processes, which have a different focus.

% The QCD NNLO studies
\begin{figure}
    \centering
    \includegraphics[width=0.47\textwidth]{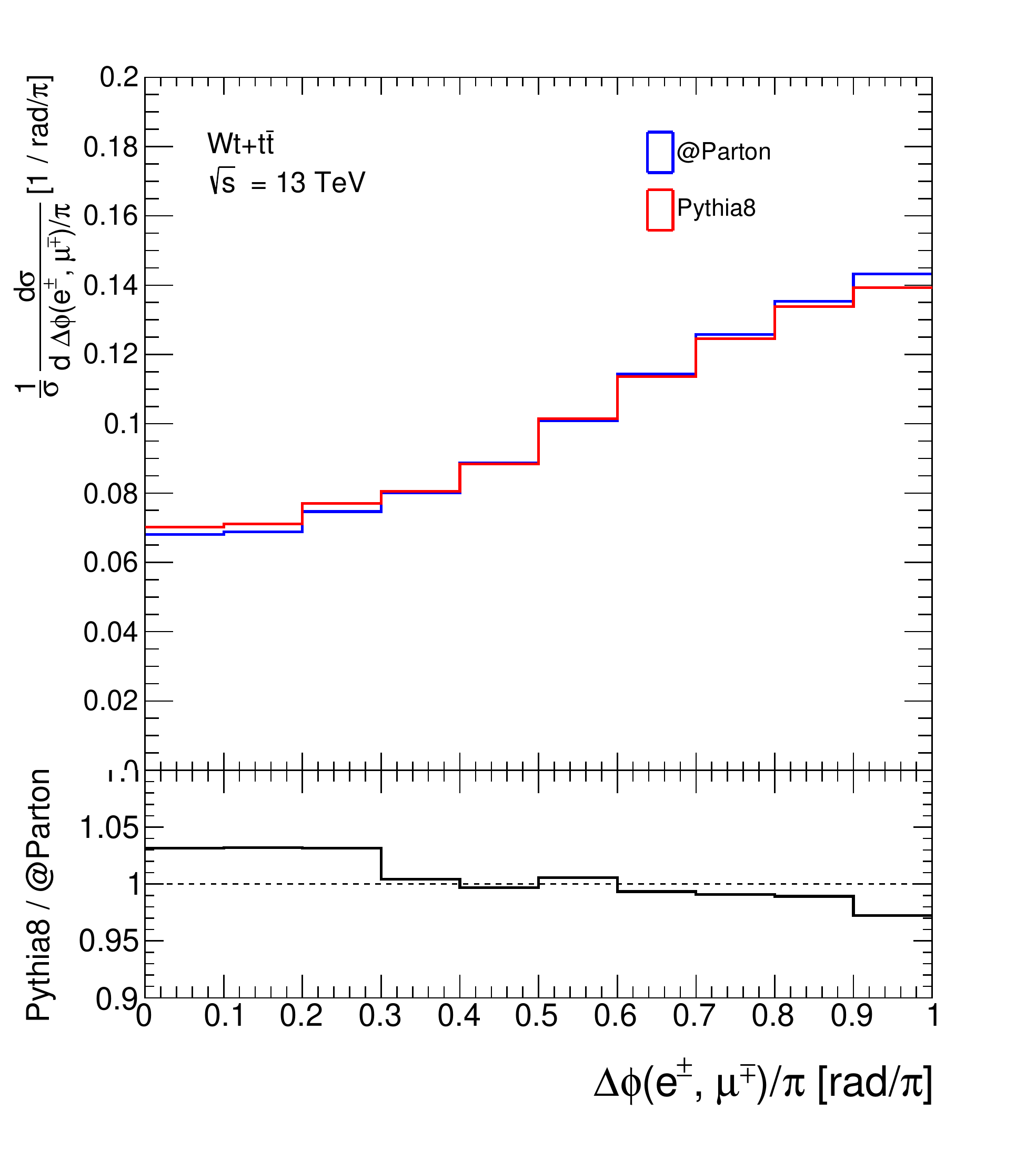}
    \includegraphics[width=0.47\textwidth]{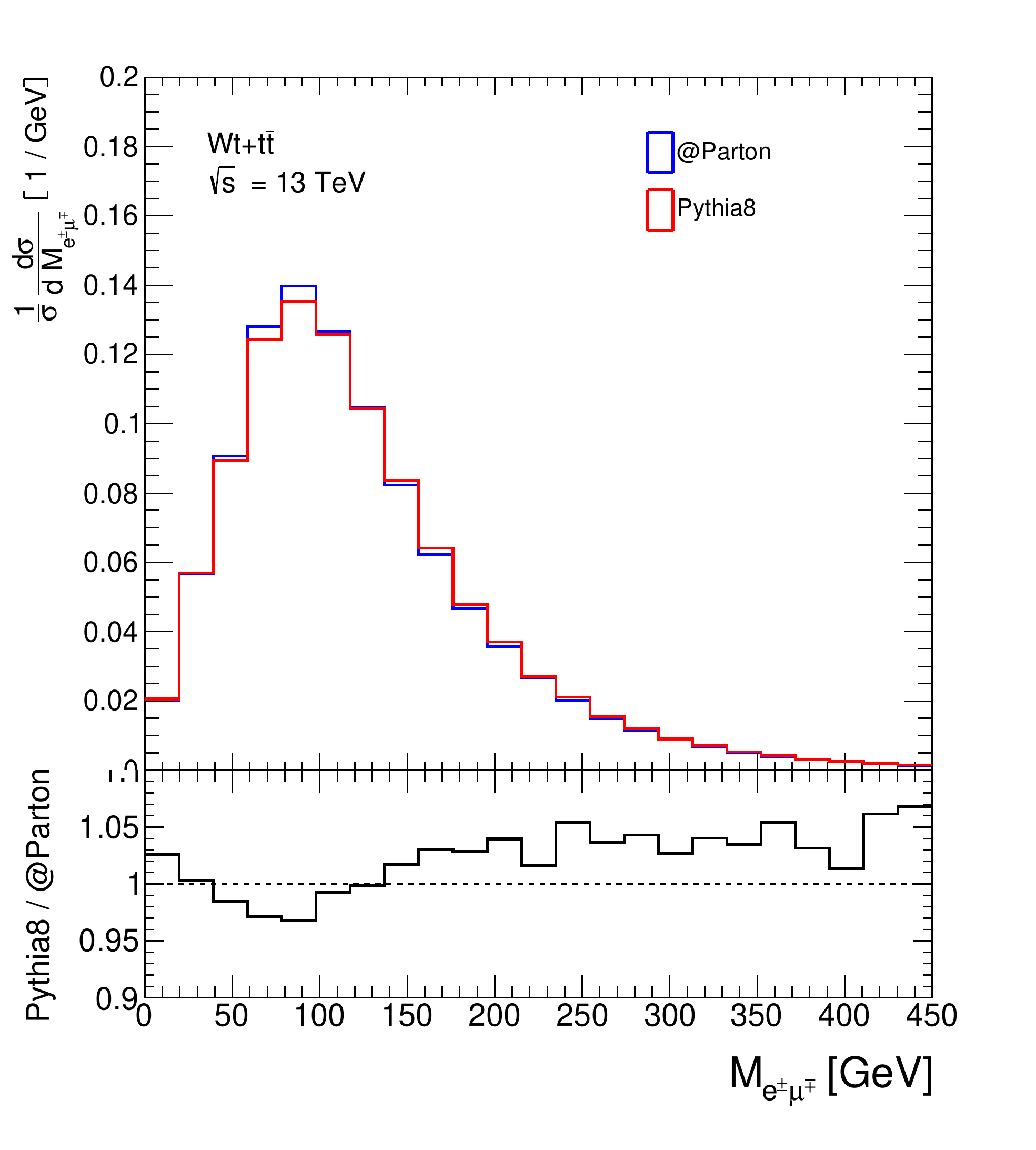}
    \caption{Impact of additional gluon radiation on the QCD NLO description of $t\overline{t}$ and $tW$ processes (see text). The left and right plots correspond to the di-lepton azimuthal angle difference and the invariant mass, respectively. }
    \label{fig:PL_VS_gluon}
\end{figure}

Another important aspect of the current investigation pertains to the impact of higher order corrections beyond QCD NLO corrections. The Matrix elements implemented in the MC used here include QCD NLO corrections. However, these are interfaced to a Parton Shower, which provides additional gluon radiation. While this setup by no means is to compete with the accuracy of a complete calculation up to QCD NNLO, one can investigate the impact of additional gluon emissions on relevant leptonic observables. In order to estimate this effect, leptonic variables obtained by means of the MC with matrix elements at QCD NLO without a parton shower, are compared to those obtained after the shower is implemented. Results are obtained using the \texttt{POWHEG} package. \Cref{fig:PL_VS_gluon} displays the comparison, where the histogram denoted by "Parton" corresponds to the generation without Parton Shower, and the one denoted by "Pythia" corresponds to the same matrix elements but with the Parton Shower. The emission of additional gluons flattens out the di-lepton azimuthal angle distribution, where configurations with small angles become favoured with respect to back-to-back configurations. This effect is qualitatively consistent with that obtained with the complete QCD NNLO corrections reported in Ref~\cite{Behring:2019iiv}. The corresponding effect on the di-lepton invariant mass is not reported in Ref~\cite{Behring:2019iiv}. The plot on the right in \Cref{fig:PL_VS_gluon} displays the effect on the di-lepton invariant mass, where the yield decreases for $m_{\ell\ell}<120$~GeV and increases for greater values. The flattening out of the azimuthal angle distribution occurs at the cost of enhancing the invariant mass. This speaks to the correlation between these two relevant leptonic observables as a result of real emissions. This correlation is also verified here by comparing results obtained with QCD NLO and NNLO corrections to the process $q\overline{q}\rightarrow W^+W^-$~\cite{Hamilton:2016bfu,Re:2018vac}: the depletion of back-to-back configurations takes place at the cost of enhancing the invariant mass.

This feature leads to an important observation: while QCD NNLO corrections improve the description of \dphill, they degrade the description of $m_{\ell\ell}$. This is driven by the differences in  correlation between \dphill and $m_{\ell\ell}$ displayed by QCD corrections and the data. Whereas the QCD corrections correlate small \dphill with large $m_{\ell\ell}$, the discrepancy between data and MC happens at small \dphill and small $m_{\ell\ell}$.

\section{Studies on \texorpdfstring{$WZ$}{WZ} theory uncertainties\label{app:wz_theory}}

\begin{figure}
    \centering
    \includegraphics[width=\textwidth]{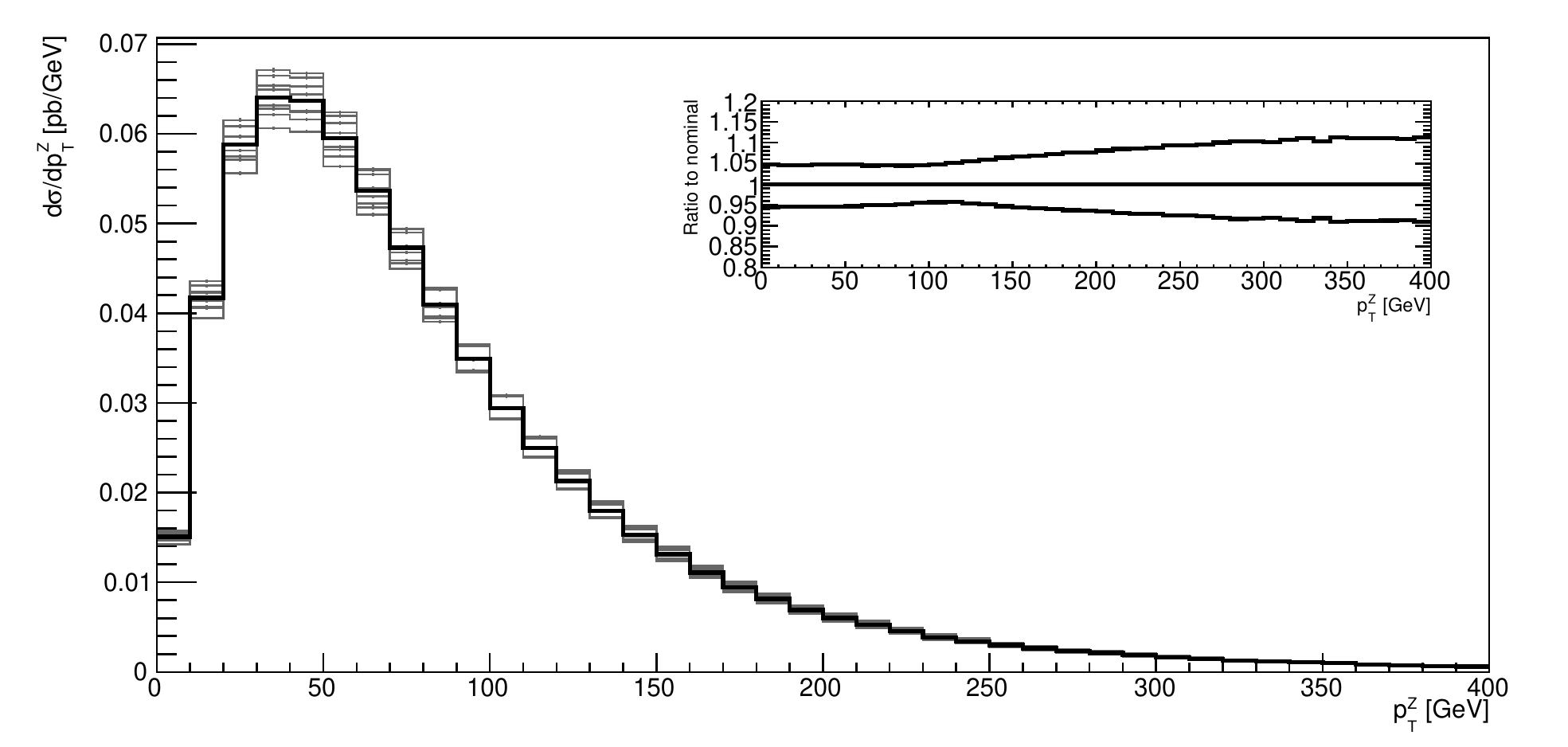}
    \caption{The effects of scale variations in the differential cross section of the SM $WZ$ process as a function of the $Z$ \pT.
    Here, \texttt{aMC@NLO} and \texttt{Pythia 8} were used to generate the events.
    The thick black line represents the spectrum at the nominal scale, and each grey line is a variation of the scale.
    The insert shows the maximum and minimum relative deviations for all scale variations.}
    \label{fig:wz_scales}
\end{figure}

In order to strengthen the statements made about the SM production of $WZ$, several theoretical studies were performed.
A variation of the dynamical renormalisation and factorisation scales (\mur and \muf, respectively) was probed at NLO using \texttt{aMC@NLO} and \texttt{Pythia 8}.
The central dynamic scale is calculated on an event-by-event basis, and is defined as half of the scalar sum of transverse momenta for the final state particles ($\HT/2$).
For the study, events were generated and passed through the \texttt{Delphes 3} fast simulation package.
The CMS Run 2 $WZ$ event selection for the measurement discussed in \Cref{sec:wz_results} was used for the purposes of this study.
The effect of varying \mur and \muf by a factor of 0.5, 1, and 2 was determined by finding the maximum and minimum deviations from the nominal case in terms of the differential cross section as a function of the $Z$ \pT, as shown in \Cref{fig:wz_scales}.
As can be seen in the spectrum, the associated scale uncertainty is of the order of 5\% in the region of $\pT < 100$~GeV, and grows to around 10\% at high \pT.
This check was also performed with the \texttt{POWHEG} event generator and the \texttt{Herwig 7} parton shower, the results of which are compatible with \Cref{fig:wz_scales}.

\begin{figure}
    \centering
    \includegraphics[width=\textwidth]{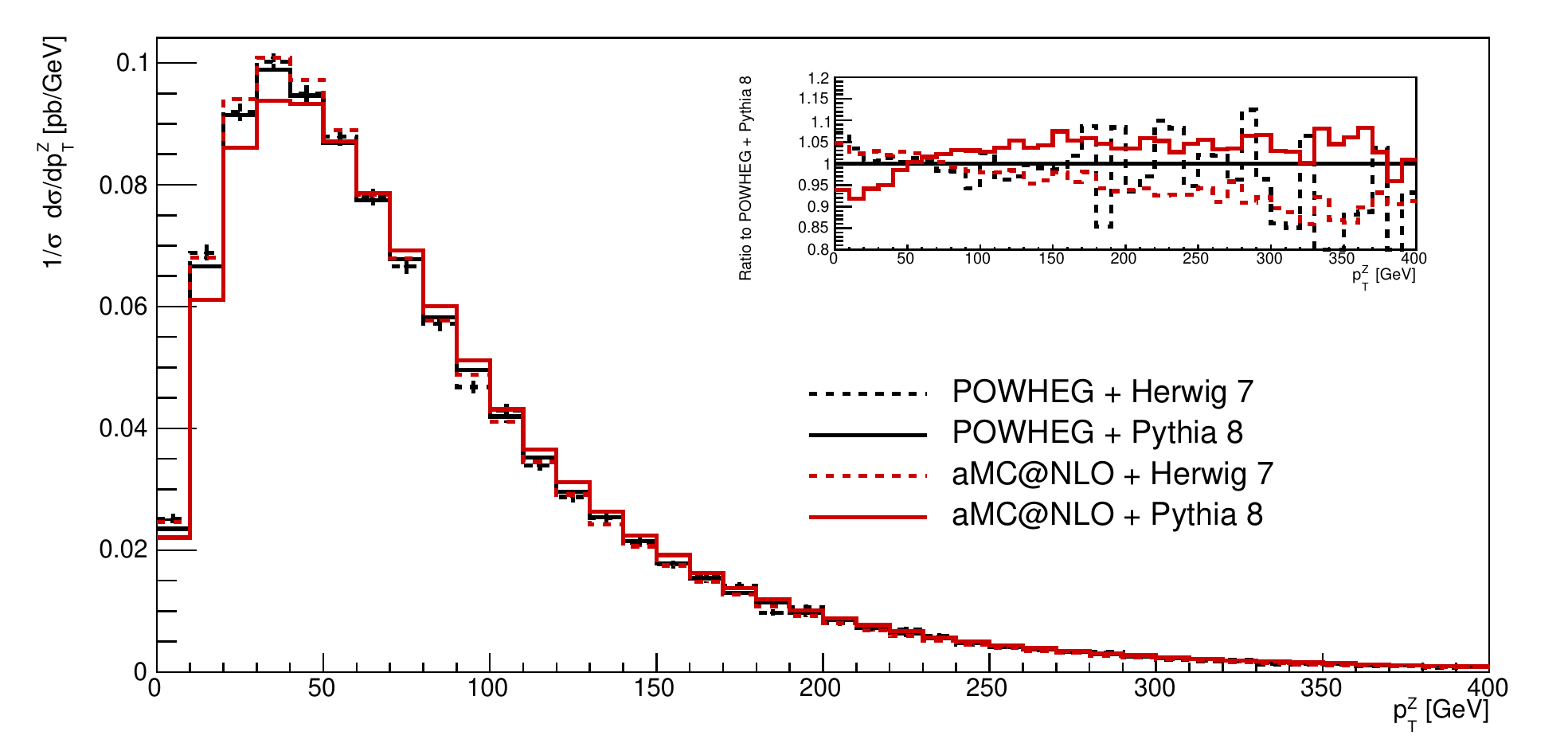}
    
    \includegraphics[width=\textwidth]{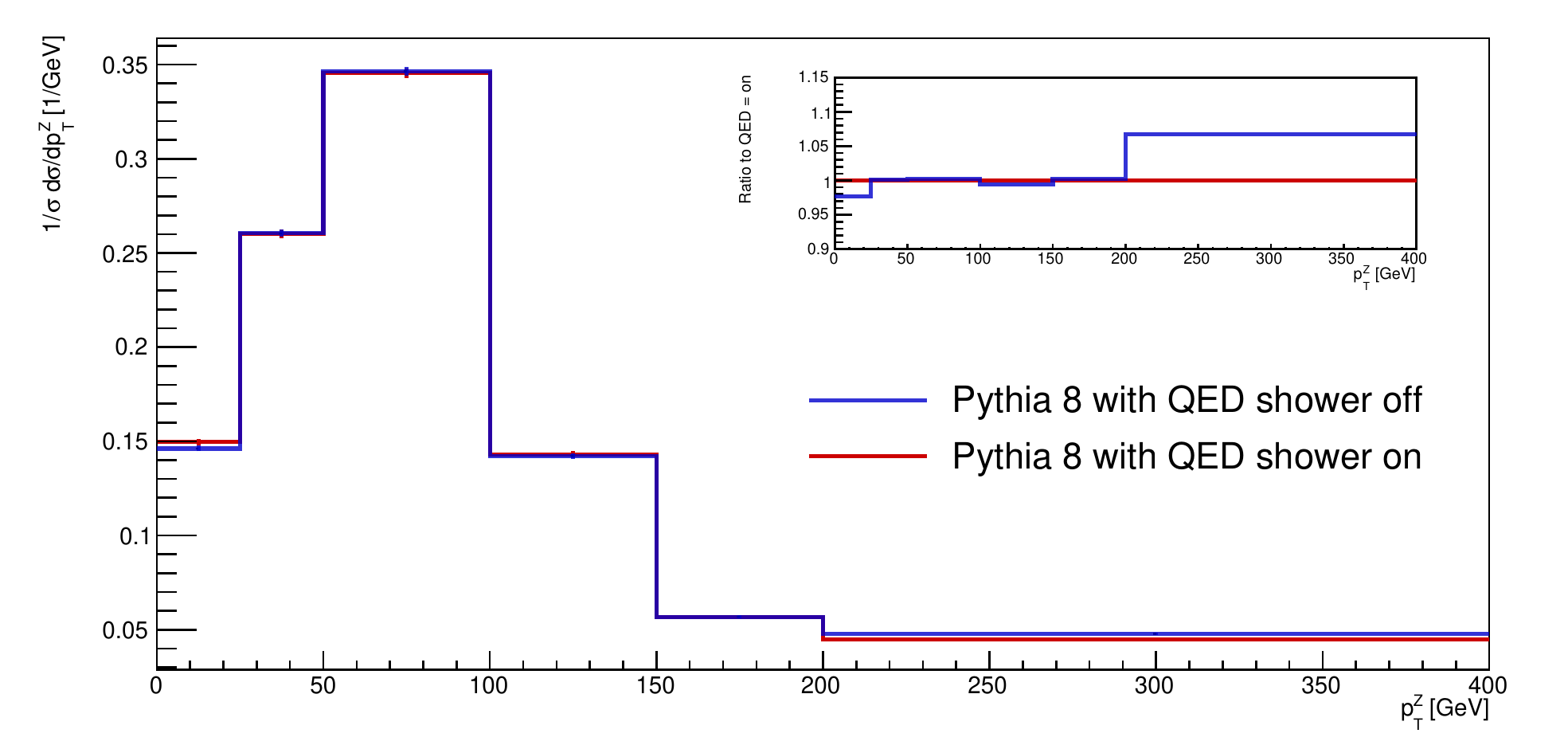}
    \caption{Normalised $Z$ \pT spectra for studying shape variations in SM $WZ$ production.
    Above, the effect of changing the event generator and parton shower program. 
    Below, the approximate effect of EW corrections determined by switching photonic emissions on and off in \texttt{Pythia 8}.}
    \label{fig:wz_shapediffs}
\end{figure}

Using the same event generation procedure as above, effects on the shape of the $Z$ \pT spectrum were also studied in two contexts.
Firstly, the difference in the $Z$ \pT spectrum for different combinations of event generators and parton shower programs were determined using the central scale cases for both \texttt{aMC@NLO} as well as \texttt{POWHEG}, and the nominal \texttt{Pythia 8} and \texttt{Herwig 7} setups.
The normalised $Z$ \pT spectrum for the four different combinations can be seen in the top of \Cref{fig:wz_shapediffs}.
Here, the nominal case can be thought of as \texttt{POWHEG} + \texttt{Pythia 8}, since both ATLAS and CMS use this for their measurements.
The biggest deviation from the nominal is therefore \texttt{aMC@NLO} + \texttt{Pythia 8}, the reason of which is unclear.
However, it can be said that changing from \texttt{POWHEG} to \texttt{aMC@NLO} would have accentuated the apparent excess at low $Z$ \pT by an unrealistic amount.
Therefore, it was decided that \texttt{POWHEG} + \texttt{Pythia 8} models the SM $WZ$ production the best, and so this prediction was used in the fits.

NNLO QCD corrections are available, where their impact has been calculated for the total cross-section and differential cross-sections pertaining to leptonic decays~\cite{Grazzini:2016swo,Grazzini:2017ckn}. The impact on the total cross-section is not of relevance here. The NNLO corrections with respect to the calculation at NLO on the $Z$ \pT differential distribution lie in the range of 5-10\%. The correction decreases as the $Z$ \pT decreases. The shape uncertainties calculated with NLO matrix elements here cover well those reported in Ref.~\cite{Grazzini:2017ckn} for the region of interest. 

Finally, due to the fact that the MC event generators which are used in the measurements do not account for higher order EW corrections, an estimate on the effect of the real radiative corrections to the $Z$ \pT was made by altering the \texttt{Pythia 8} parton shower.
SM $WZ$ events were generated at LO in QCD using \texttt{Pythia 8}, in order to separate the EW corrections from the QCD corrections.
The $Z$ \pT spectrum with and without photonic emissions in the shower can be seen in the bottom of \Cref{fig:wz_shapediffs}.
It can be seen that the most significant correction to the \pT in terms of real EW corrections comes at high \pT, which is far from the region in which the BSM model in this article predicts a signal.
In this region, the current statistical imprecision of the data dominates over the potential for the $Z$ \pT spectrum to be significantly altered by EW corrections.
Of course, virtual EW corrections should also play a role, however these become significant for large values of $Z$ \pT.

Soft gluon effects have been studied in the context of the production of $VV, V=W,Z$.  ~\cite{Grazzini:2005vw,Frederix:2008vb,Wang:2013qua,Becher:2019bnm}. Whereas theoretical errors can be sizeable at low values of $q_T$, or the transverse momentum of the $VV$ system, the region of interest in the \pT of the $Z$ boson is not particularly affected by uncertainties. 

\section{Impact of systematic uncertainties\label{app:syst}}

\begin{figure}
    \centering
    \begin{subfigure}[b]{0.315\textwidth}
    \includegraphics[width=\textwidth]{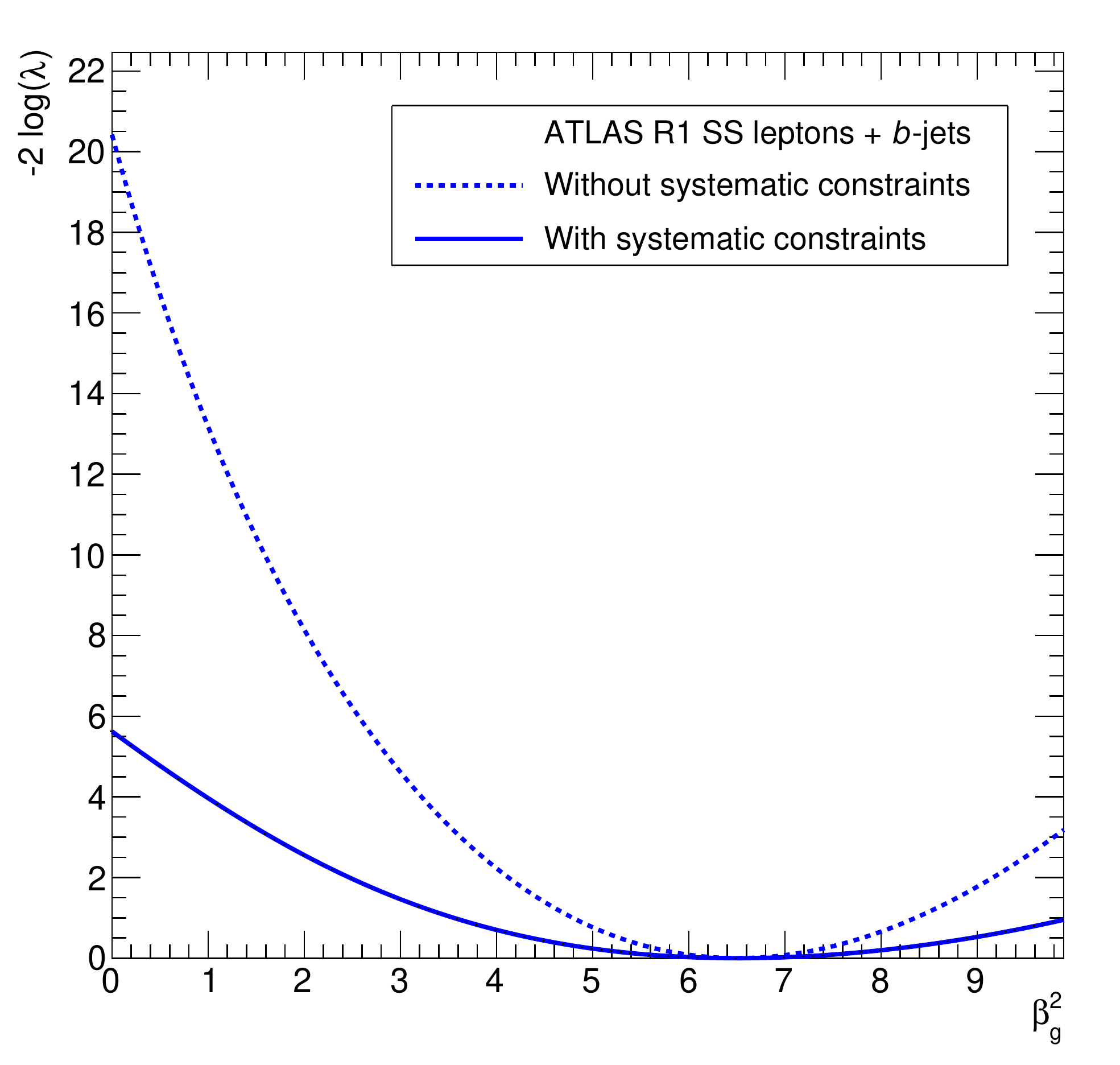}
    \caption{ATLAS-EXOT-2013-16}
    \end{subfigure}
    ~
    \begin{subfigure}[b]{0.315\textwidth}
    \includegraphics[width=\textwidth]{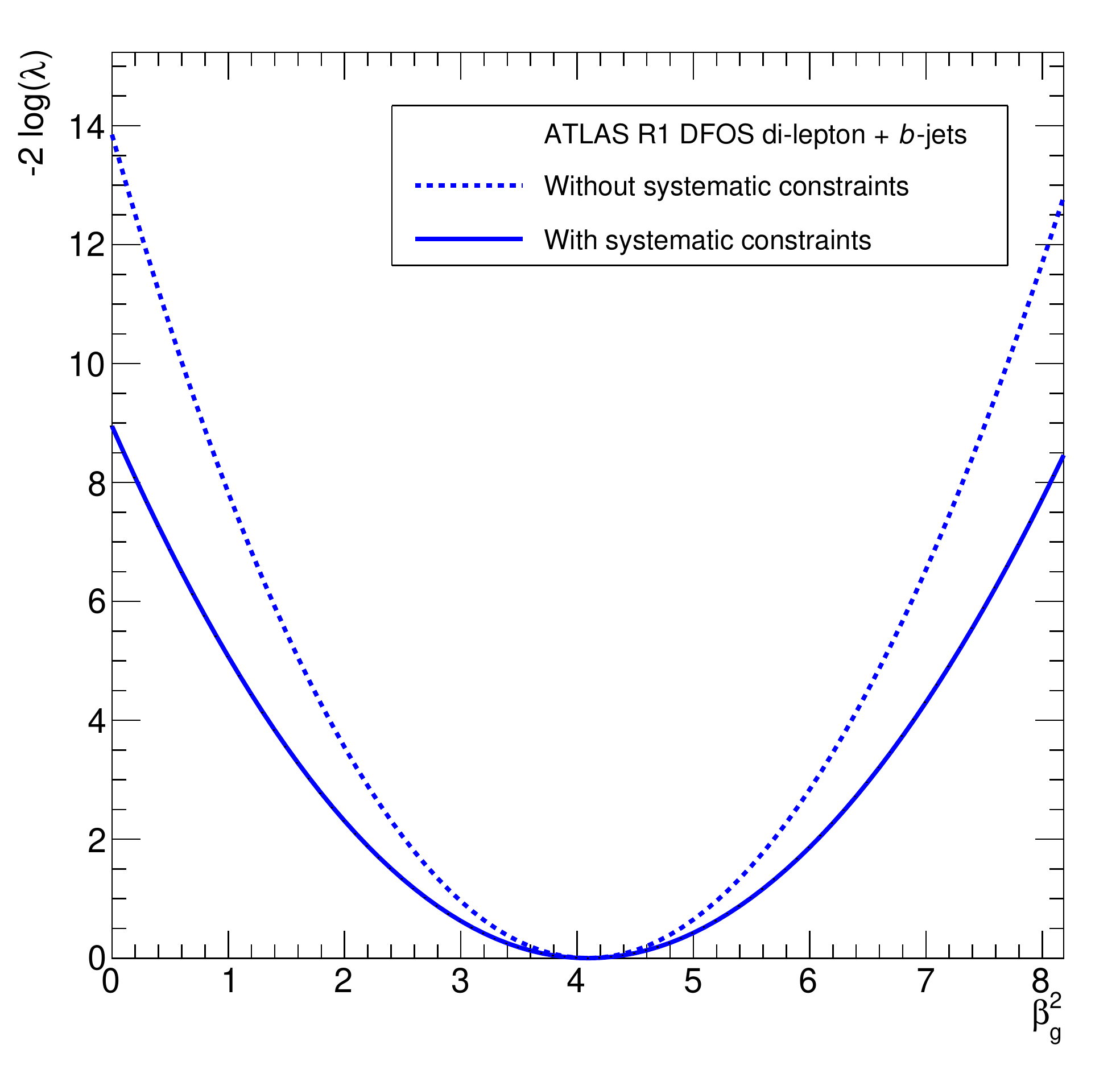}
    \caption{ATLAS-TOPQ-2015-02}
    \end{subfigure}
    ~
    \begin{subfigure}[b]{0.315\textwidth}
    \includegraphics[width=\textwidth]{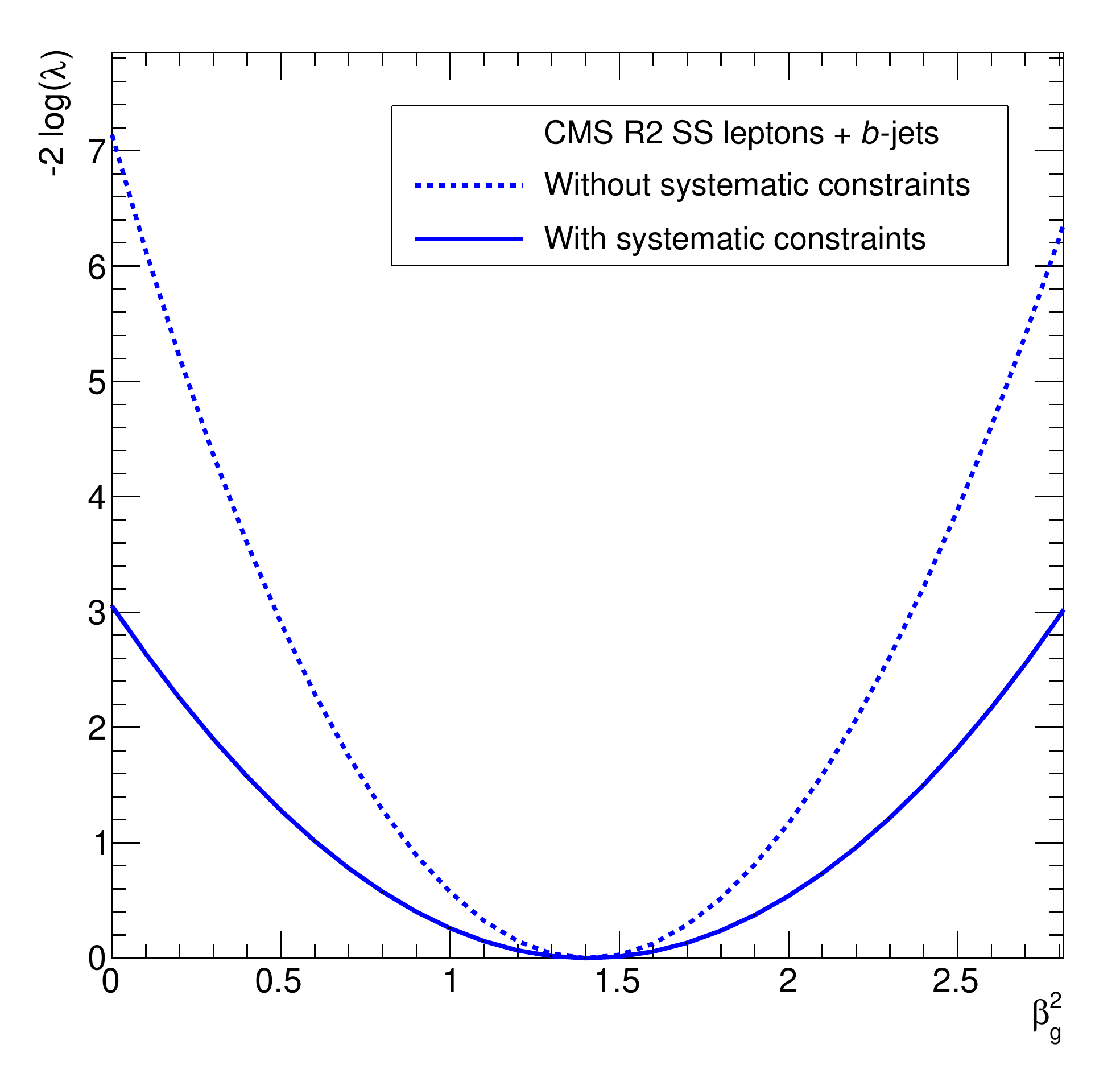}
    \caption{CMS-PAS-HIG-17-005}
    \end{subfigure}
    
    \vspace{50pt}
    
    \begin{subfigure}[b]{0.315\textwidth}
    \includegraphics[width=\textwidth]{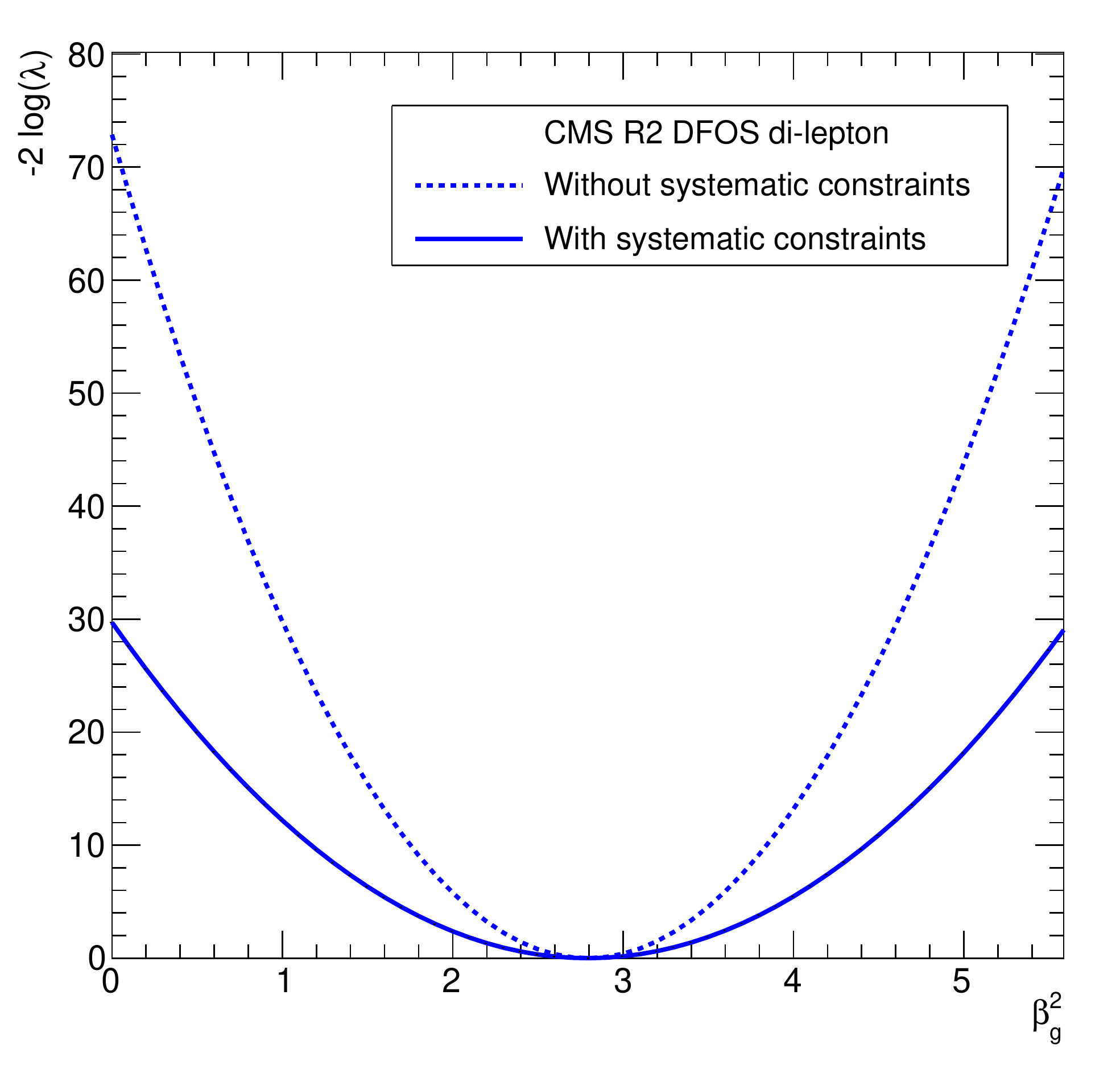}
    \caption{CMS-TOP-17-018}
    \end{subfigure}
    ~
    \begin{subfigure}[b]{0.315\textwidth}
    \includegraphics[width=\textwidth]{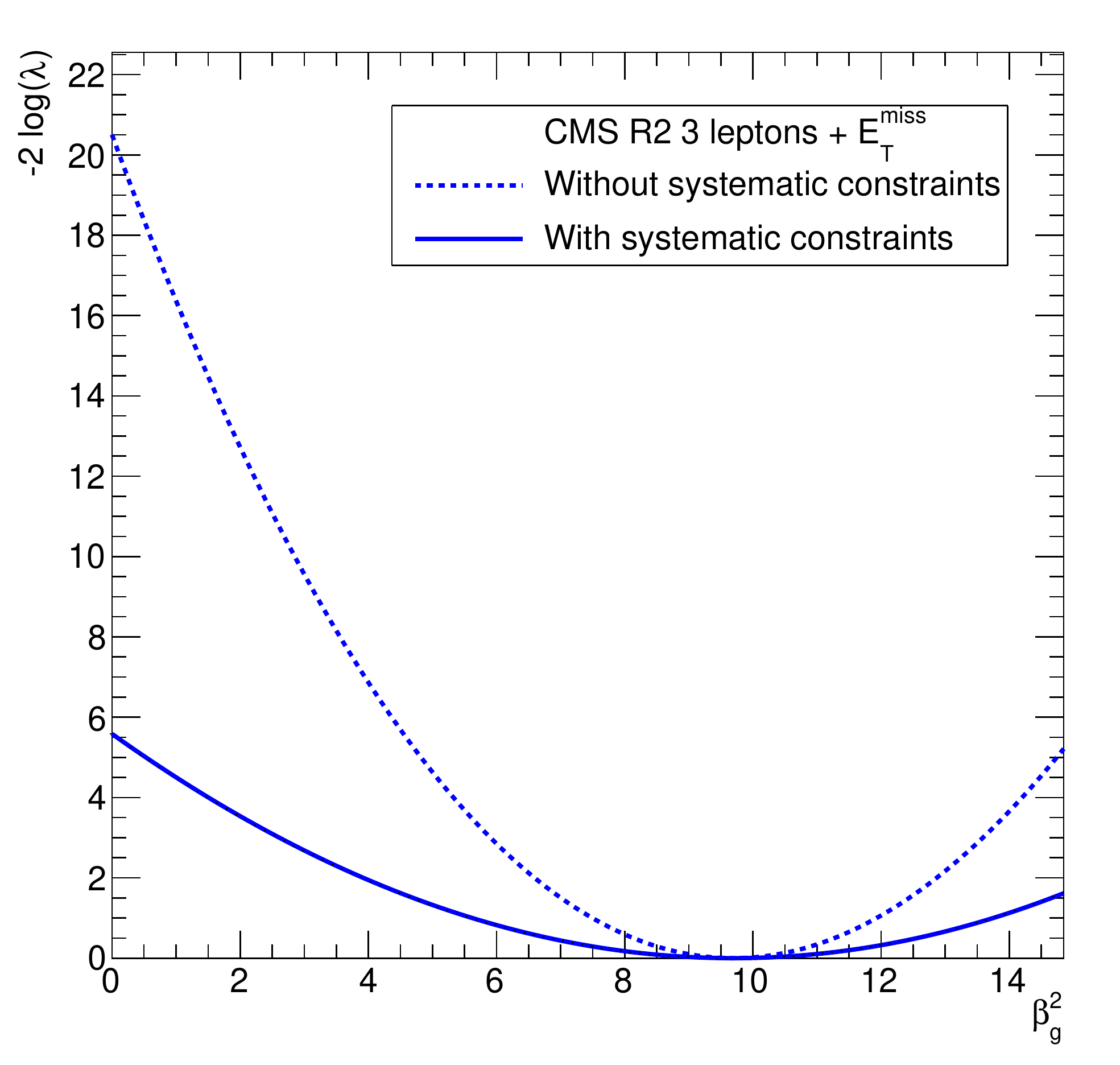}
    \caption{CMS-PAS-SMP-18-002}
    \end{subfigure}
    ~
    \begin{subfigure}[b]{0.315\textwidth}
    \includegraphics[width=\textwidth]{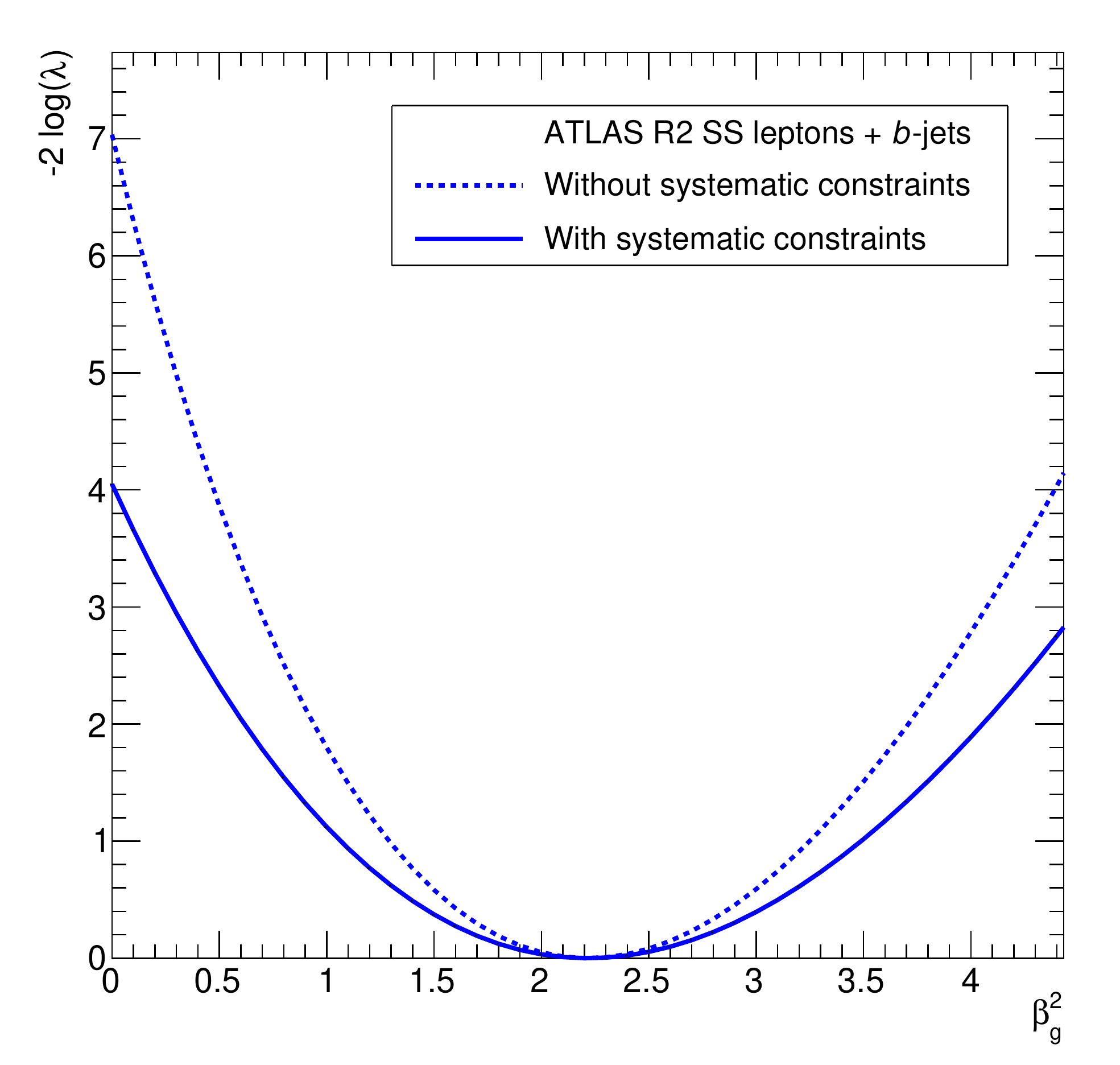}
    \caption{ATLAS-EXOT-2016-16}
    \end{subfigure}
    
    \vspace{50pt}
    
    \begin{subfigure}[b]{0.315\textwidth}
    \includegraphics[width=\textwidth]{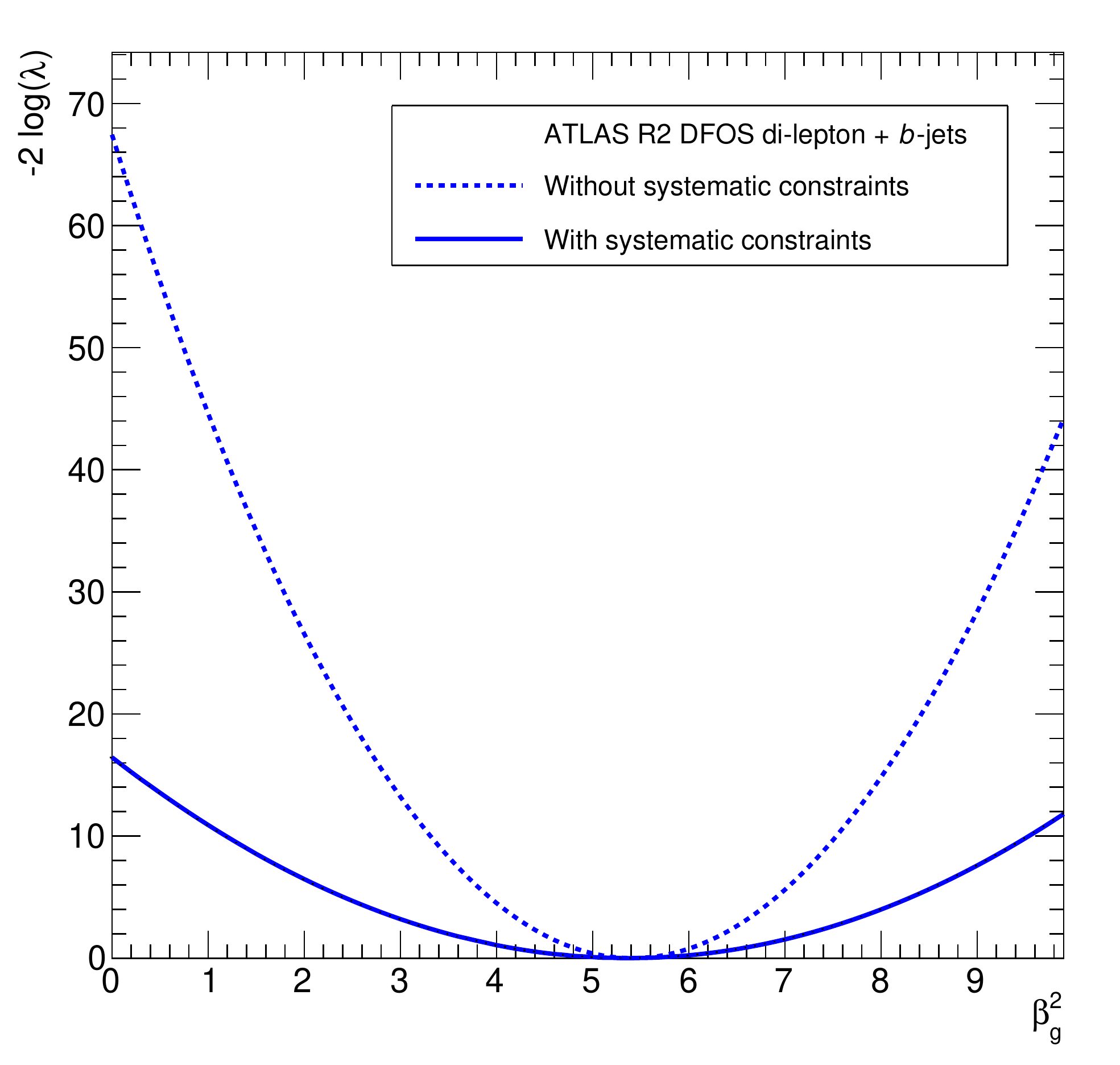}
    \caption{ATLAS-CONF-2018-027}
    \end{subfigure}
    ~
    \begin{subfigure}[b]{0.315\textwidth}
    \includegraphics[width=\textwidth]{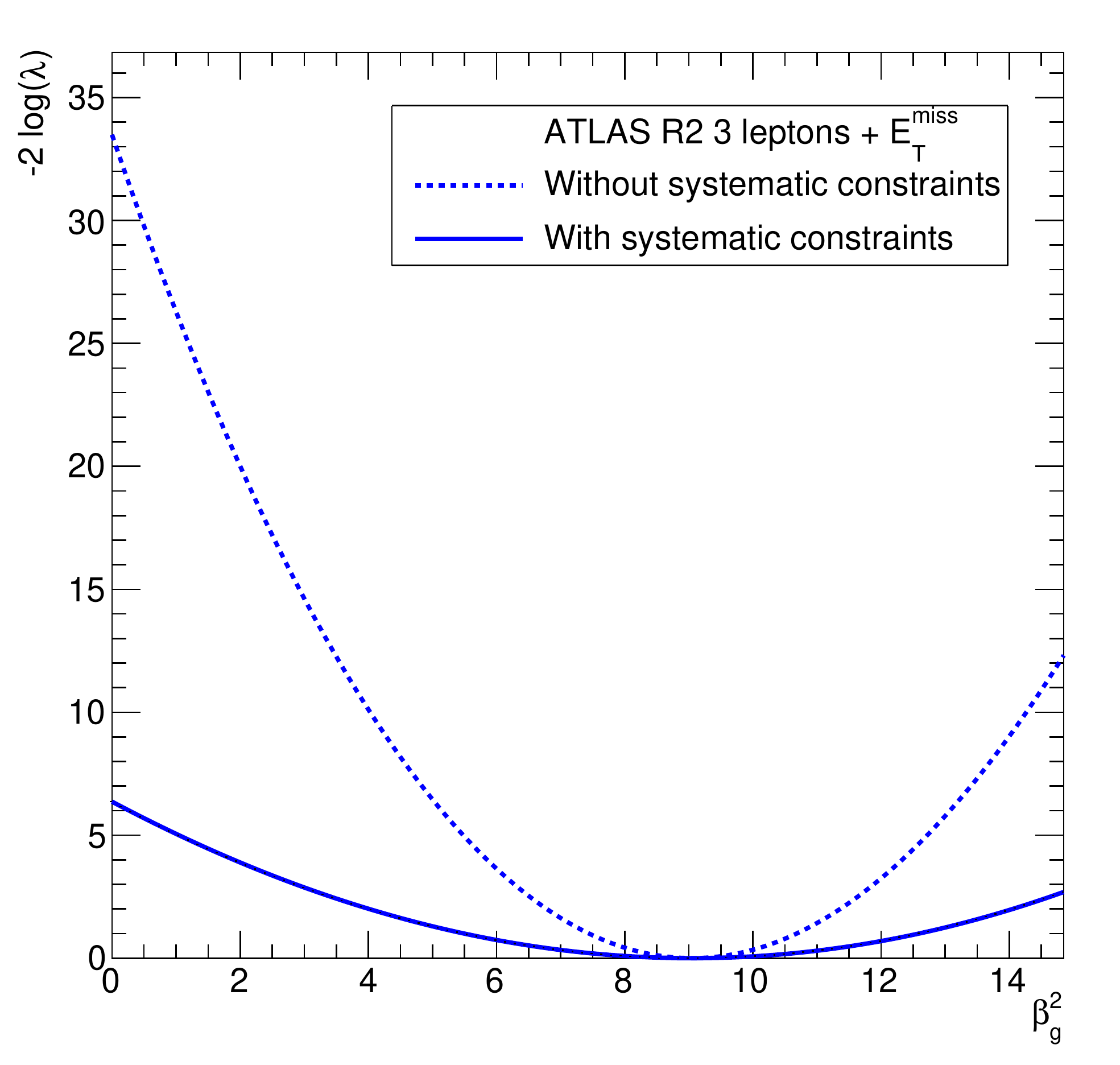}
    \caption{ATLAS-CONF-2018-034}
    \end{subfigure}
    ~
    \begin{subfigure}[b]{0.315\textwidth}
    \includegraphics[width=\textwidth]{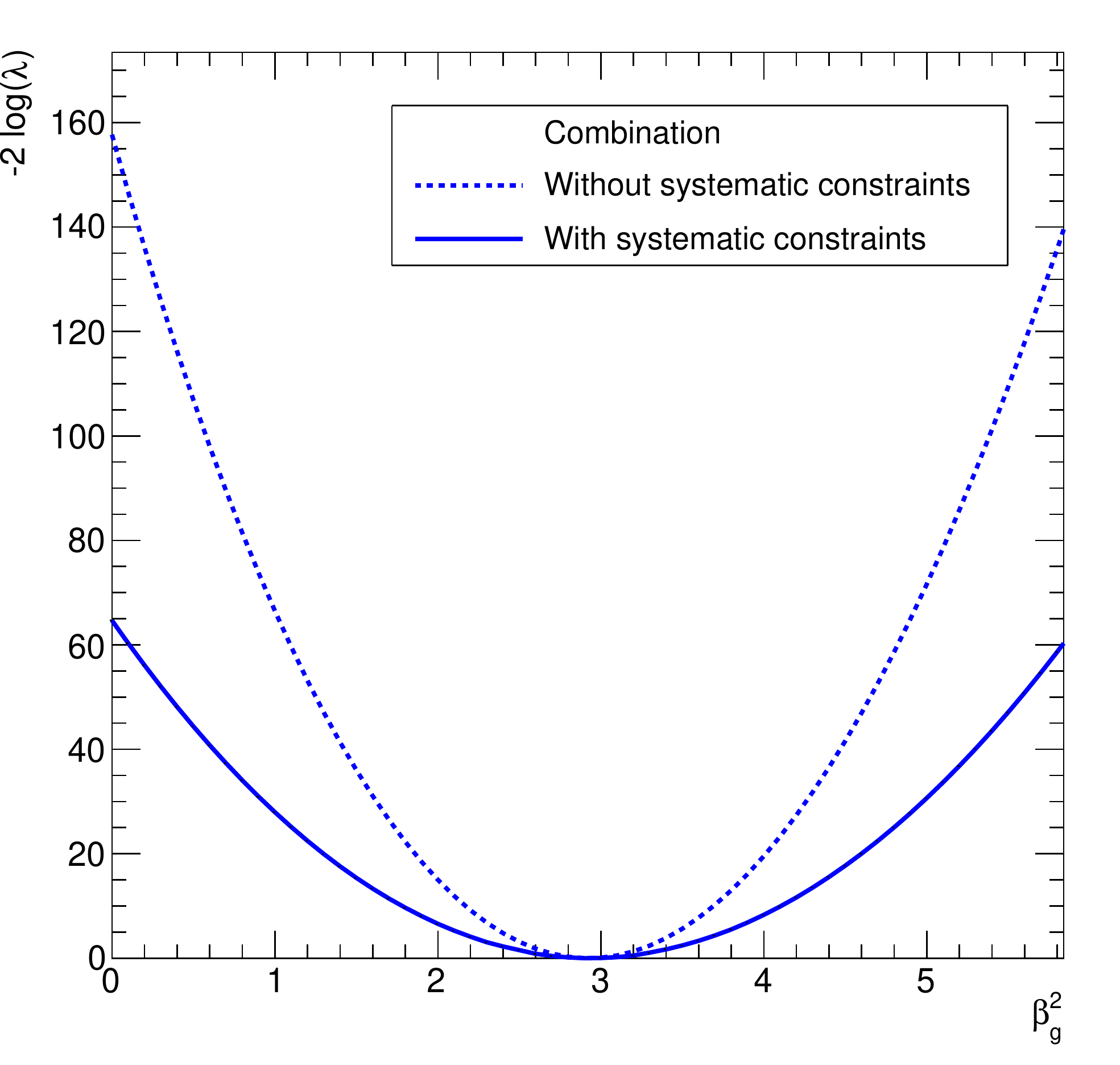}
    \caption{Combination}
    \end{subfigure}
    \caption{The profile likelihood ratio overlaid with the NLL for each of the fit results performed in this article.
    These figures provide a good reference for how the systematic uncertainties affect the significance of the obtained fit results.}
    \label{fig:systematics}
\end{figure}

Due to the high statistical precision of many of the results studied in this article, it is important to understand the effect of systematic uncertainties on each of the fit results.
In \Cref{fig:systematics}, the profile likelihood ratio is shown overlaid with the corresponding NLL, as a function of \bgs.
Since the NLL does not contain information about the overall constraints of the systematic uncertainties, the comparison of the two is a good measure by which we can understand the effects of systematic uncertainties on the fit results.

\bibliographystyle{JHEP.bst}
\bibliography{references.bib}

\end{document}